\documentclass[%
 reprint,
 superscriptaddress,
 amsmath,amssymb,
 prl,
floatfix,
]{revtex4-2}
\usepackage{graphicx}
\usepackage[caption=false]{subfig}
\usepackage{dcolumn}
\usepackage{bm}
\usepackage{hyperref}

\usepackage[separate-uncertainty=true,list-units=repeat,multi-part-units=brackets]{siunitx}
[separate-uncertainty=true,list-units=repeat,multi-part-units=brackets]

\DeclareSIUnit\bar{bar}

\begin{document}

\title{Resonant excitation of plasma waves in a plasma channel}

\author{A. J. Ross}
\email{aimee.ross@physics.ox.ac.uk}
\affiliation{John Adams Institute for Accelerator Science and Department of Physics, University of Oxford, Denys Wilkinson Building, Keble Road, Oxford OX1 3RH, United Kingdom}%

\author{J. Chappell}
\affiliation{John Adams Institute for Accelerator Science and Department of Physics, University of Oxford, Denys Wilkinson Building, Keble Road, Oxford OX1 3RH, United Kingdom}%

\author{J. J. van de Wetering}
\affiliation{John Adams Institute for Accelerator Science and Department of Physics, University of Oxford, Denys Wilkinson Building, Keble Road, Oxford OX1 3RH, United Kingdom}%

\author{J. Cowley}
\affiliation{John Adams Institute for Accelerator Science and Department of Physics, University of Oxford, Denys Wilkinson Building, Keble Road, Oxford OX1 3RH, United Kingdom}%

\author{E. Archer}
\affiliation{John Adams Institute for Accelerator Science and Department of Physics, University of Oxford, Denys Wilkinson Building, Keble Road, Oxford OX1 3RH, United Kingdom}%

\author{N. Bourgeois}
\affiliation{Central Laser Facility, STFC Rutherford Appleton Laboratory, Didcot OX11 0QX, United Kingdom}%

\author{L. Corner}
\affiliation{Cockcroft Institute for Accelerator Science and Technology, School of Engineering, The Quadrangle, University of Liverpool, Brownlow Hill, Liverpool L69 3GH, United Kingdom}

\author{D. R. Emerson}
\affiliation{Scientific Computing Department, STFC Daresbury Laboratory, Warrington WA4 4AD, United Kingdom}

\author{L. Feder}
\affiliation{John Adams Institute for Accelerator Science and Department of Physics, University of Oxford, Denys Wilkinson Building, Keble Road, Oxford OX1 3RH, United Kingdom}%

\author{X. J. Gu}
\affiliation{Scientific Computing Department, STFC Daresbury Laboratory, Warrington WA4 4AD, United Kingdom}

\author{O. Jakobsson}
\affiliation{John Adams Institute for Accelerator Science and Department of Physics, University of Oxford, Denys Wilkinson Building, Keble Road, Oxford OX1 3RH, United Kingdom}%

\author{H. Jones}
\altaffiliation[Now at ]{Deutsches Elektronen-Synchrotron DESY, Notkestr. 85, 22607 Hamburg, Germany}
\affiliation{Cockcroft Institute for Accelerator Science and Technology, School of Engineering, The Quadrangle, University of Liverpool, Brownlow Hill, Liverpool L69 3GH, United Kingdom}

\author{A. Picksley}
\altaffiliation[Now at ]{Lawrence Berkeley National Laboratory, Berkeley, California 94720, USA}
\affiliation{John Adams Institute for Accelerator Science and Department of Physics, University of Oxford, Denys Wilkinson Building, Keble Road, Oxford OX1 3RH, United Kingdom}%

\author{L. Reid}
\affiliation{Cockcroft Institute for Accelerator Science and Technology, School of Engineering, The Quadrangle, University of Liverpool, Brownlow Hill, Liverpool L69 3GH, United Kingdom}

\author{W. Wang}
\affiliation{John Adams Institute for Accelerator Science and Department of Physics, University of Oxford, Denys Wilkinson Building, Keble Road, Oxford OX1 3RH, United Kingdom}%

\author{R. Walczak}
\affiliation{John Adams Institute for Accelerator Science and Department of Physics, University of Oxford, Denys Wilkinson Building, Keble Road, Oxford OX1 3RH, United Kingdom}%
\affiliation{Somerville College, Woodstock Road, Oxford OX2 6HD, United  Kingdom}%

\author{S. M. Hooker}
\affiliation{John Adams Institute for Accelerator Science and Department of Physics, University of Oxford, Denys Wilkinson Building, Keble Road, Oxford OX1 3RH, United Kingdom}%

\date{\today}

\begin{abstract}
We demonstrate resonant excitation of a plasma wave by a train of short laser pulses guided in a pre-formed plasma channel, for parameters relevant to a plasma-modulated plasma accelerator (P-MoPA). We show experimentally that a train of $N \approx 10$ short pulses, of total energy  $\sim \SI{1}{J}$, can be guided through \SI{110}{mm} long plasma channels with on-axis densities in the range \SI{E17} - \SI{E18}{cm^{-3}}. The spectrum of the transmitted train is found to be strongly red-shifted when the plasma period is tuned to the intra-train pulse spacing. Numerical simulations are found to be in excellent agreement with the measurements and indicate that the resonantly excited plasma waves have an amplitude in the range 3 - \SI{10}{GV.m^{-1}}, corresponding to an accelerator stage energy gain of order \SI{1}{GeV}. 
\end{abstract}

\maketitle

In the laser wakefield accelerator (LWFA) \cite{tajima79}, a short laser pulse propagating through a plasma excites a trailing Langmuir wave, within which the generated electric fields can be of the order $E_\mathrm{wb} = m_\text{e} c \omega_p / e$, where $\omega_p = (n_\text{e} e^2 / m_\text{e} \epsilon_0)^{1/2}$ is the plasma frequency, and $n_\text{e}$ is the electron density. For electron densities of interest $E_\mathrm{wb} \sim \SI{100}{GV.m^{-1}}$, some three orders of magnitude greater than is possible in a conventional accelerator. Considerable progress has been made, including, for example, the acceleration of electrons to energies in the GeV range in centimetre-scale accelerator stages~\cite{leemans2006, Kneip.2009, Wang.2013hwz, leemans2014, Shin.2018, gonsalves2019, miao2022, Oubrerie2022LSA, Picksley2023}, and the application of LWFAs to driving compact light sources \cite{Corde.2013, Albert.2016}. Recently, free-electron laser gain was demonstrated using laser-accelerated electrons~\cite{Wang.2021, Labat.2022nd8}.

To drive a large amplitude Langmuir (or `plasma') wave, the duration $\tau_L$  of the laser pulse must satisfy $\tau_L \lesssim T_p /2$, where $T_p = 2 \pi / \omega_p$ is the plasma period, corresponding to $\tau_L \lesssim \SI{100}{fs}$ for plasma densities of interest. As a consequence, recent experimental work has been dominated by the use of high energy (joule-scale) chirped-pulse-amplification \cite{strickland1985} Ti:sapphire lasers. However, this laser material has a high quantum defect (34\%)~\cite{Siders.2019} which limits the pulse repetition rate of high-energy systems to $f_\mathrm{rep} \ll \SI{1}{kHz}$.

An alternative method for driving the plasma wave is to resonantly excite it with a train of low-energy pulses (or a single long, modulated pulse) in which the pulse spacing (or modulation) is matched to $T_p$. An example of this approach is the plasma beat-wave accelerator (PBWA) \cite{tajima79, Joshi.1984, clayton1993, tochitsky2004}, in which two long pulses of frequencies $\omega_1$ and $\omega_2 = \omega_1 + \omega_p$ are combined to form a  pulse modulated at $\omega_p$. Beat-wave acceleration of electrons to energies in the \SI{10}{MeV} range has been reported; of particular relevance to the present work is that by Tochitsky \textit{et al.}~\cite{tochitsky2004}, who exploited ponderomotive self-guiding over \SI{3}{cm} to accelerate electrons to \SI{38}{MeV} at a gradient of $\sim \SI{1}{GV.m^{-1}}$.

\begin{figure*}[tb]
    \centering
    \includegraphics[width = 0.98\textwidth]{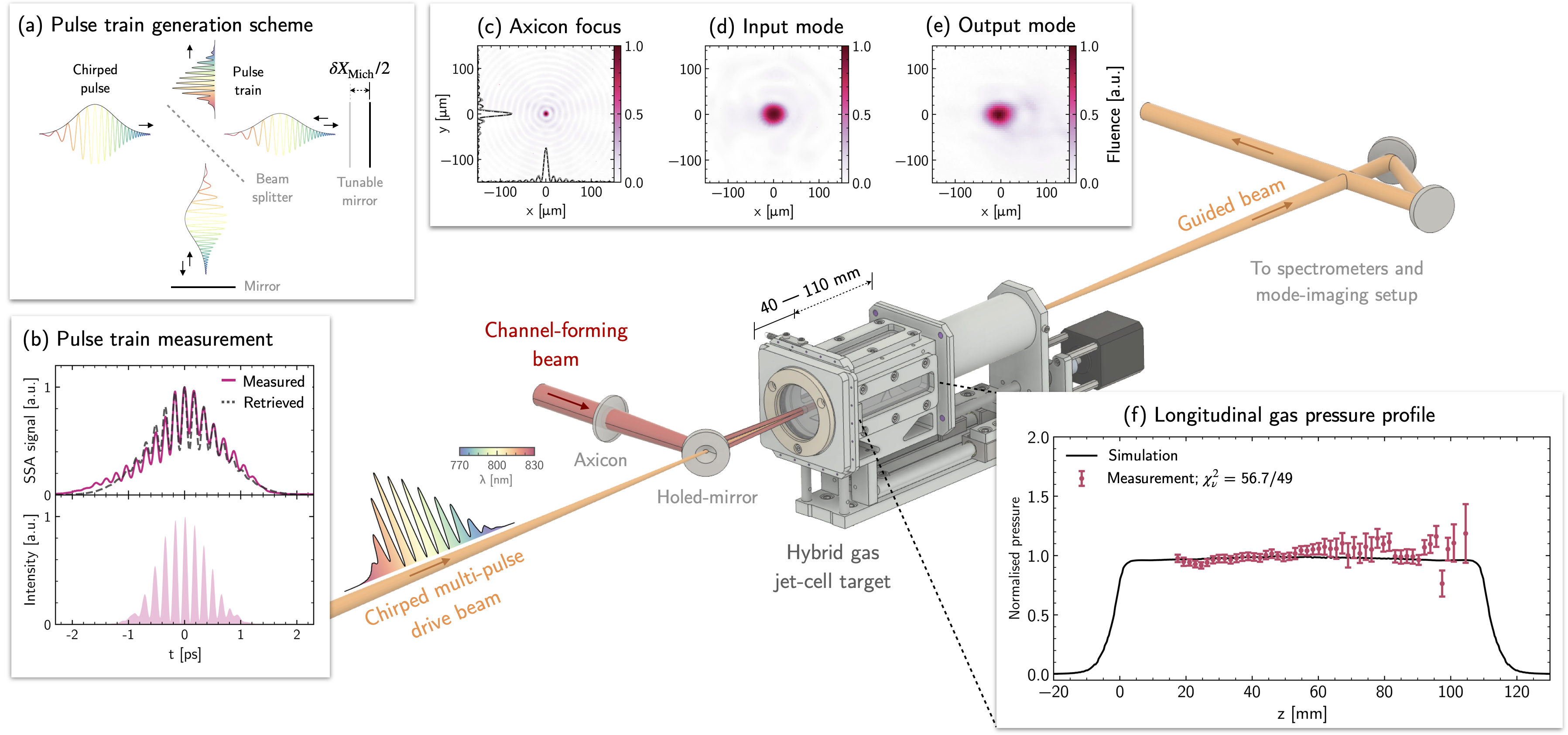}
    \caption{Sketch of the experimental layout. (a) Illustration of the pulse train generation scheme. (b) Example single-shot autocorrelator (SSA) measurement for the $\tau = \SI{170(2)}{fs}$ pulse train. Upper: comparison between the measured (pink) and retrieved (grey, dashed) SSA signal. Lower: retrieved pulse train intensity profile. (c) Measured axicon focus. (d) Example input mode of the focused multi-pulse drive beam. (e) Example guided mode at the channel exit. All focal spot images are normalized to their maximum. (f) Comparison between the measured and simulated longitudinal gas pressure profile~\cite{Supp_mat}.}
    \label{fig:Fig1}
\end{figure*}

Interest in resonant wakefield excitation has revived \cite{Hooker_2014} with the development of novel laser technologies, such as thin-disk lasers that can generate joule-scale pulses at $f_\mathrm{rep}$ in the kilohertz range, with high ($\gtrsim 10\%$) wall-plug efficiency~\cite{Wang.2020}. The picosecond-duration pulses provided by these systems are too long to drive a plasma wave directly, and a second laser frequency separated by $\omega_p$ is not currently available to drive a PBWA. A potential solution is the plasma-modulated plasma accelerator (P-MoPA)~\cite{jakobssonPMoPA}, which comprises three stages: (i) a modulator, in which a long ($\sim \SI{1}{ps}$), high-energy ($\gtrsim \SI{1}{J}$) laser pulse is spectrally modulated by the low amplitude plasma wave driven by a short ($\lesssim \SI{100}{fs}$), low-energy ($\lesssim \SI{100}{mJ}$) `seed' laser pulse as they co-propagate in a plasma channel of on-axis density $n_\mathrm{e,0}$; (ii) a dispersive optical system that converts the spectral modulation to a train of short pulses spaced by $T_{p,0} = 2 \pi \sqrt{m_\text{e} \epsilon_0 / n_\mathrm{e,0} e^2}$; (iii) an accelerator stage, also of on-axis density $n_\mathrm{e,0}$, within which the pulse train resonantly drives a large amplitude plasma wave. Numerical simulations~\cite{jakobssonPMoPA} show that a \SI{1.7}{J}, \SI{1}{ps} driver, with a \SI{140}{mJ}, \SI{40}{fs} seed, could accelerate electrons to energies of \SI{0.65}{GeV} in a \SI{100}{mm}-long plasma channel with $n_\mathrm{e,0} = \SI{2.5E17}{cm^{-3}}$. 

In this Letter we investigate experimentally the accelerator stage of a P-MoPA. We demonstrate guiding of a train of $N \approx 10$ short pulses, with a total energy of the order \SI{1}{J} through \SI{110}{mm} long plasma channels, equivalent to 14 Rayleigh ranges, with $n_\mathrm{e,0}$ in the range \SI{E17} - \SI{E18}{cm^{-3}}. Resonant excitation of a plasma wave within the channel is evidenced by the observation of strong red-shifting of the spectrum of the transmitted pulse train when $T_{p,0}$ was tuned to the pulse spacing in the train. The results are found to be in excellent agreement with numerical simulations, which show that wake amplitudes in the range \SIrange{3}{10}{GV.m^{-1}} were achieved, corresponding to an accelerator stage energy gain of the order \SI{1}{GeV}.

Figure~\ref{fig:Fig1} shows schematically the arrangement employed for these experiments, undertaken with the Astra-Gemini TA3 Ti:sapphire laser at the Rutherford Appleton Laboratory. This laser provides two synchronized beams, here denoted the `drive' and `channel-forming' beams, each of central wavelength $\lambda_0 =  $ \SI{800}{nm} with transform-limited full-width at half-maximum (FWHM) duration of \SI{31}{fs}. In order to mimic the pulse train employed in the P-MoPA scheme, single laser pulses were converted to a train of short pulses using a Michelson interferometer, as sketched in Fig.~\ref{fig:Fig1}(a) and described previously~\cite{Shalloo2016, cowley2017}. The temporal intensity profile of the generated pulse train, shown in Fig.~\ref{fig:Fig1}(b), was determined from single-shot measurements of the spectrum and autocorrelation of the train (see Supplemental Material~\cite{Supp_mat} for further details).

The gas target used in this work was a cell-jet hybrid~\cite{Aniculaesei2018, Picksley2023}, with hydrogen gas pulsed into the target via a solenoid valve and two transducers measuring the pressure on-shot. The laser pulses were coupled into, and out of, the target via a pair of \SI{3}{mm} radius coaxial pinholes mounted on: (i) the front of the target; and (ii) a motorized plunger that could be moved to adjust the target length $L$. A relative RMS pressure variation along the laser propagation axis of $\SI{4.1}{\%}$ was measured~\cite{Supp_mat}, as shown in Fig.~\ref{fig:Fig1}(e). 

\begin{figure}[t]
	\centering
	\includegraphics[width = \linewidth]{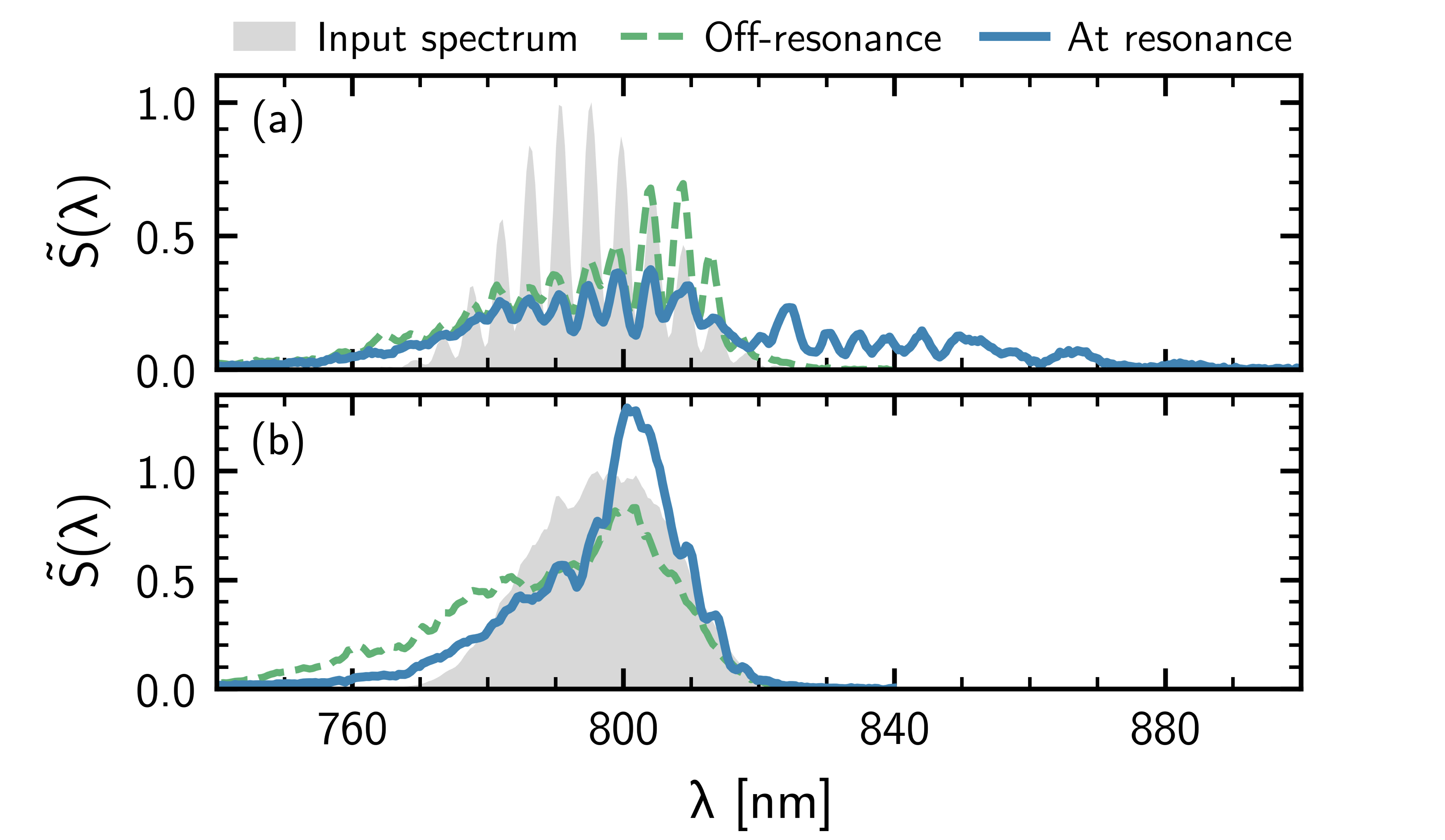}
	\caption{Comparison of the photon-normalized spectra, $\tilde{S}(\lambda)$, of the input pulses (grey) and those transmitted through a \SI{110}{mm}-long HOFI channel for: (a) a pulse train with $\tau = \SI{170}{fs}$ and $E_\mathrm{train} = \SI{2.5(0.5)}{J}$; (b) an unmodulated pulse with FWHM duration $\sim \SI{1}{ps}$ and $E =  \SI{2.7(0.5)}{J}$. $\tilde{S}(\lambda)$ is shown near the resonance condition of the pulse train [blue; $n_\mathrm{e, res} =$ \SI{4.3(0.3)e17}{cm^{-3}}] and for an off-resonant density [green, dashed; $n_\mathrm{e,0} =$ \SI{1.4(0.3)e17}{cm^{-3}}]. The photon-normalized spectra have been scaled to a maximum value of unity for the input pulse.
 }
	\label{fig:Fig2}
\end{figure}

A hydrodynamic optical-field-ionized (HOFI) channel~\cite{shalloo2018,shalloo2019} was formed in the target by focusing the channel-forming pulse, of energy $\sim \SI{100}{mJ}$ and FWHM pulse duration \SI{80}{fs}, with an axicon lens of base angle \ang{3.6}. The transverse intensity profile of the beam produced by the axicon had a central maximum of FWHM spot size \SI{9.8(0.1)}{\micro \metre}, as shown in Fig.~\ref{fig:Fig1}(c).

The pulse train, of total on-target energy $E_\mathrm{train} = \SI{2.5(0.5)}{J}$, was focused by an off-axis $f/40$ paraboloid to the target entrance. The transverse intensity profile of the focused beam [see Fig.~\ref{fig:Fig1}(d)] was found to have a $1/\mathrm{e}^2$ intensity radius of \SI{45.5(3.4)}{\micro \metre}, a Rayleigh range of  $z_R =$ \SI{7.9 \pm 0.7}{\milli \metre}, and to contain \SI{64.9(1.5)}{\%} of its energy within its FWHM. The delay between the arrival of the channel-forming and drive beams was set to $t_d =$ \SI{3.5}{\nano \second}. After leaving the plasma channel, the energy of the drive beam was reduced, and the beam re-imaged onto a 16-bit camera and a fibre-coupled spectrometer. An example guided mode is shown in Fig.~\ref{fig:Fig1}(e).

\begin{figure}[t]
    \centering
    \includegraphics[width = \linewidth]{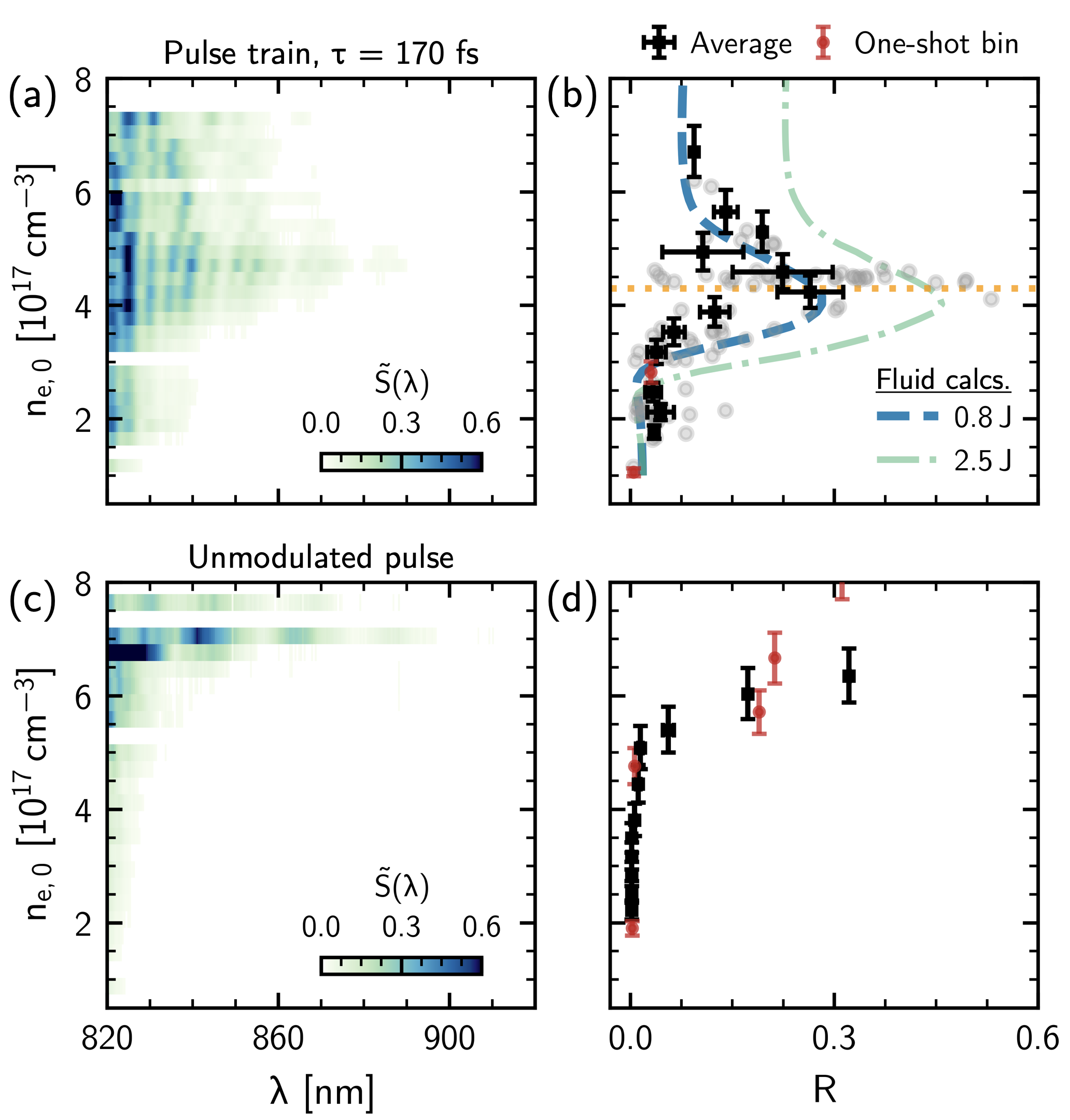}
    \caption{Density dependence of $\tilde{S}(\lambda)$ for: (a,b) a pulse train with $\tau = \SI{170}{fs}$, $E_\mathrm{train} = \SI{2.5(0.5)}{J}$; and (c,d) an unmodulated $\sim \SI{1}{ps}$, \SI{2.7(0.5)}{J}  pulse. (a) and (c): $\tilde{S}(\lambda)$, averaged in electron density bins of width $\Delta n_\text{e,0} = \SI{0.24e17}{cm^{-3}}$. (b) and (d): mean values of $R$ [black squares] weighted by energy transmission; red circles indicate bins containing data from only a single shot. The $n_\mathrm{e,0}$-error bars are a combination of the uncertainties in the measured pressure and the on-axis density calibration. The $R$-error bars represent the standard error on the mean. In (b), individual data points are plotted (grey circles) and the orange dotted line represents $n_{\text{e,res}}(\tau = \SI{170}{fs})$. Overlaid are the results of the fluid calculations for $E_\mathrm{train} =  \SI{2.5}{J}$ (green) and $E_\mathrm{train} = \SI{0.8}{J}$ (blue). 
    }
    \label{fig:Fig3}
\end{figure}

The excitation of plasma waves by the drive pulse was detected through changes in its spectrum~\cite{EsareyFreqShifts1990}. The spectra presented in Figs.~\ref{fig:Fig2} and~\ref{fig:Fig3} are photon-normalized, defined as $\tilde{S}(\lambda) = \lambda S_\mathrm{meas}(\lambda) / \int_0^\infty \lambda S_\mathrm{meas}(\lambda) \mathrm{d} \lambda$, where $S_\mathrm{meas}(\lambda)$ is the measured spectrum. Figure~\ref{fig:Fig2}(a) shows $\tilde{S}(\lambda)$ for an incident pulse train with $E_\mathrm{train} = \SI{2.5(0.5)}{J}$ and $\tau =\SI{170}{fs}$, at on-axis densities approximately equal to, and one third of, the resonant value, $n_\mathrm{e,res} \approx \SI{4.3e17}{cm^{-3}}$. As expected, the input spectrum of the pulse train is modulated by the Michelson interferometer to yield $N \approx 10$, uniformly-spaced peaks. For the off-resonant density, the spectrum of the transmitted train is similar to that of the incident pulse, with some blue-shifting apparent in the region $\lambda \lesssim \SI{780}{nm}$, likely caused by ionization of the neutral gas collar \cite{chofi2020, feder2020} surrounding the HOFI channel and of the gas plumes that extend beyond the target. In contrast, at the resonant density, considerable red-shifting is observed, extending the bandwidth of the input beam by more than \SI{40}{nm} on the long wavelength side. The new red-shifted light beyond \SI{820}{nm} is seen to consist of a series of peaks~\cite{jakobssonPMoPA}; these arise from spectral modulation of the laser pulse by the wakefield, which generates copies of the input spectrum shifted by $\pm m \omega_p$ for integer $m$. The peaks on the blue side of the spectrum are not visible in Fig.~\ref{fig:Fig2}, likely due to the additional blue-shift from ionisation. We note that blue-shifting would have predominantly occurred for the first few pulses in the train, and, since the pulse train was negatively chirped, their initial spectra were on the blue side of the mean wavelength.

The density-dependent red-shift seen in Fig.\ \ref{fig:Fig2}(a) strongly indicates resonant plasma wave excitation in the plasma channel. To confirm this, we also measured the transmitted spectra for a temporally-smooth $\sim \SI{1}{ps}$ drive pulse of similar energy at on-axis densities matching those in Fig.\ \ref{fig:Fig2}(a). As shown in Fig.~\ref{fig:Fig2}(b), in this case no red-shift was observed, and the spectra were similar for both densities and were dominated by blue-shift of similar magnitude to that observed in Fig.~\ref{fig:Fig2}(a). 

Figure~\ref{fig:Fig3} shows the variation with on-axis plasma density of the transmitted spectra when the drive was well-guided~\cite{Supp_mat} by the plasma channel. To quantify the red-shift we define the red-shift metric $R= \sum_{\lambda_\mathrm{min}}^{\infty} \tilde{S}(\lambda)$, where $\lambda_\mathrm{min}$ is the longest wavelength in the input spectrum above the noise level. It is evident from Fig.\ \ref{fig:Fig3}(a,b) that the spectra of the pulse train driver exhibit a pronounced red-shift for densities in the range $n_\mathrm{e,0} = \SIrange{4}{5}{\times 10^{17} cm^{-3}}$, which agrees with the expected resonance density of $n_\mathrm{e,res} =\SI{4.3e17}{cm^{-3}}$. For a train of $N$ identical laser pulses, the full-width of the resonance peak is expected \cite{cowley2017} to be $\delta n_\mathrm{e,0}/n_\mathrm{e,res} \approx 8/(3N)$, corresponding to $\delta n_\mathrm{e,0} \approx$ \SI{1.2e17}{cm^{-3}} --- in good agreement with the measured FWHM in $R$ of $\delta n_\mathrm{e,0} \approx \SI{1.6e17}{cm^{-3}}$. In contrast, Figs.\ \ref{fig:Fig3}(c, d) show that no resonance is observed for the unmodulated drive pulse. Significant red-shifting of the unmodulated drive pulse \emph{is} observed for $n_\mathrm{e,0} \gtrsim \SI{5.5e17}{cm^{-3}}$, likely caused by self-modulation~\cite{PhysRevLett.33.209, esarey1994selfmod, nakajima1995selfmod} of the long pulse. 

To provide further insight, we compared these measurements with the results of an in-house 2D cylindrical fluid code, benchmarked against the particle-in-cell (PIC) code WarpX \cite{WarpX} (see \cite{Supp_mat}). The calculations used the retrieved pulse train parameters and modelled the plasma channel as an ideal fully-ionized parabolic waveguide \cite{Supp_mat}. The code ignores the effects of ionization by the laser pulse, and assumes that the temporal envelope of the drive is unchanged by its interaction with the plasma. 

Figure~\ref{fig:Fig3}(b) shows the calculated $R$ for the $\tau = \SI{170}{ns}$, $E_\mathrm{train} = \SI{2.5}{J}$ pulse train in a plasma channel of length $L = \SI{110}{mm}$. It can be seen that the position and width of the calculated resonance peak agree closely with those observed in the measurements. For some shots the measured $R$ values reach the calculated curve, but in most cases they are lower. In order to understand this, the energy transmission of the train was measured as a function of the plasma channel length~\cite{Supp_mat}. For each cell length the measured energy transmission was found to vary over a wide range, owing to the large pointing jitter of the input pulse train. Shots for which the input beam was well aligned with the channel axis were found to have an input coupling of $T_0 = \SI{64(4)}{\%}$, which is consistent with $|c_0|^2 = (71 \pm 5)\%$, where $c_0$ is the calculated~\cite{Supp_mat} coupling coefficient between the transverse amplitude profile of the input beam and that of the lowest-order mode of the channel. In contrast, the coupling coefficient deduced from all guided shots is only $T_0 = \SI{32(13)}{\%}$, which reflects the additional losses arising from misalignment with respect to the channel axis. Figure~\ref{fig:Fig3}(b) shows that if the drive energy is reduced by this factor, i.e.\ to $E_\mathrm{train} = \SI{800}{\milli \joule}$, the calculated variation of $R$ with density is in excellent agreement with the averaged measurements. At the resonant density, the amplitude of the wakefield driven by the $\tau = \SI{170}{fs}$ pulse train is calculated from the fluid simulation to be \SI{10}{GV m^{-1}} (\SI{3}{GV m^{-1}}) for $E_\text{train} = \SI{2.5}{J}\; (\SI{0.8}{J})$. 

\begin{figure}[t]
    \centering
    \includegraphics[width=\linewidth]{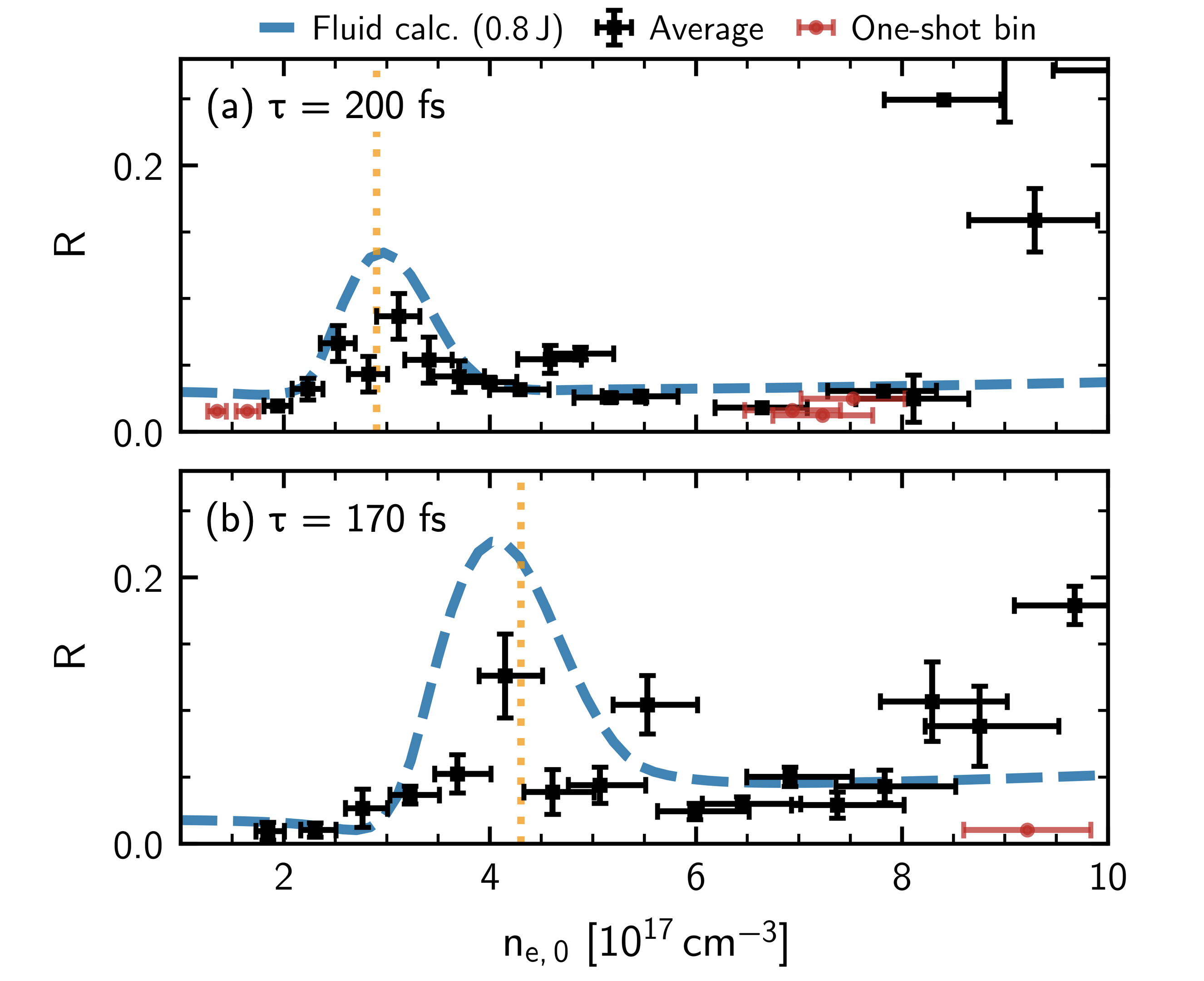}
    \caption{Variation of $R$ with on-axis density for a plasma channel of length $L =$ \SI{70}{mm} and for pulse trains of energy \SI{2.7(0.5)}{J} and pulse separation: (a) $\tau =$ \SI{200}{fs}; and (b) $\tau =$ \SI{170}{fs}. The results of the fluid calculations, assuming $E_{\text{train}} = \SI{800}{\milli \joule}$, are shown by the blue dashed lines. For each plot the expected resonant density is indicated by the orange dotted line.
    }
    \label{fig:Fig4}
\end{figure}

Further evidence of resonant wakefield excitation is shown in Fig.\ \ref{fig:Fig4}, which shows the measured and calculated variation of $R$ with on-axis density for pulse trains with $\tau = $ \SI{200}{fs} and \SI{170}{fs}. In this case, $E_\mathrm{train} = \SI{2.5(0.5)}{J}$ and $L = \SI{70}{mm}$. It can be seen that, for both pulse trains, the position, width, and magnitude of the measured variation of $R$ agree well with the calculation assuming $E_\mathrm{train} = \SI{0.8}{J}$. At higher densities, $n_\mathrm{e,0} \gtrsim$ \SI{7e17}{cm^{-3}}, red-shifting arising from self-modulation is again observed.

It has been previously shown that HOFI~\cite{shalloo2018,shalloo2019} channels achieve higher energy transmission when the wings of the laser pulse have sufficient intensity to ionize the neutral gas collar to form a conditioned~\cite{picksley2020, miao2020} HOFI channel. PIC simulations~\cite{Supp_mat} of the present experiment indicate that the leading three pulses in the train conditioned the HOFI channel, allowing later pulses in the train to be guided with low losses. We note that conditioning of the channel could also be achieved by employing a separate, short pulse immediately ahead of the pulse train~\cite{chofi2020}; the required energy of the conditioning pulse is $\sim \SI{7}{mJ}$ per cm of channel, i.e.\ only 3\% of the drive energy in the present experiment.

In summary we have demonstrated guiding of a train of $N \approx 10$ short pulses, with a total pulse train energy of the order \SI{1}{J} through \SI{110}{mm} long plasma channels with on-axis densities in the range \SI{E17} - \SI{E18}{cm^{-3}}. The spectra of the transmitted pulse trains were found to be strongly red-shifted when the plasma period was matched to the pulse spacing in the train. In contrast, no such resonance in the red-shift was observed for an unmodulated drive pulse of the same total energy and duration. Numerical simulations were found to be in excellent agreement with the measurements, and showed that, at resonance, the wake amplitude was in the range $3 - \SI{10}{GV.m^{-1}}$, corresponding to an accelerator stage energy gain of the order \SI{1}{GeV}. 

These results constitute the first demonstration of resonant excitation of a plasma wave by a train of laser pulses guided in a pre-formed plasma channel. The laser and plasma parameters employed in this work are directly relevant to the accelerator stage of the P-MoPA scheme~\cite{jakobssonPMoPA}, which offers a route to achieving kilohertz-repetition-rate, GeV-scale plasma accelerators driven by plasma modulation of joule-scale, picosecond-duration laser pulses, such as those provided by thin-disk lasers.

\acknowledgements
This work was supported by the UK Engineering and Physical Sciences Research Council (EPSRC) (Grant Nos EP/R513295/1 \& EP/V006797/1), the UK Science and Technologies Facilities Council (Grant Nos ST/P002048/1, ST/R505006/1, ST/S505833/1, ST/V001655/1, ST/V001612/1), and the Ken and Veronica Tregidgo Scholarship in Atomic and Laser Physics. This work required significant computing resources which were funded by the plasma HEC Consortium [EPSRC Grant
No. EP/R029149/1] and UKRI funding [ARCHER2 Pioneer Projects]. Computing resources were provided by ARCHER and ARCHER2 [ARCHER2 PR17125] UK supercomputers \url{http://archer.ac.uk}, \url{https://www.archer2.ac.uk}. This research used the open-source particle-in-cell code WarpX \url{https://github.com/ECP-WarpX/WarpX}, primarily funded by the US DOE Exascale Computing Project. Primary WarpX contributors are with LBNL, LLNL, CEA-LIDYL, SLAC, DESY, CERN, and TAE Technologies. We acknowledge all WarpX contributors.

Data is available from the authors upon reasonable request.

This research was funded in whole, or in part, by EPSRC and STFC, which are Plan S funders. For the purpose of Open Access, the author has applied a CC BY public copyright licence to any Author Accepted Manuscript version arising from this submission.

\bibliography{references}

\providecommand{\noopsort}[1]{}\providecommand{\singleletter}[1]{#1}%
\begin{thebibliography}{44}%
\makeatletter
\providecommand \@ifxundefined [1]{%
 \@ifx{#1\undefined}
}%
\providecommand \@ifnum [1]{%
 \ifnum #1\expandafter \@firstoftwo
 \else \expandafter \@secondoftwo
 \fi
}%
\providecommand \@ifx [1]{%
 \ifx #1\expandafter \@firstoftwo
 \else \expandafter \@secondoftwo
 \fi
}%
\providecommand \natexlab [1]{#1}%
\providecommand \enquote  [1]{``#1''}%
\providecommand \bibnamefont  [1]{#1}%
\providecommand \bibfnamefont [1]{#1}%
\providecommand \citenamefont [1]{#1}%
\providecommand \href@noop [0]{\@secondoftwo}%
\providecommand \href [0]{\begingroup \@sanitize@url \@href}%
\providecommand \@href[1]{\@@startlink{#1}\@@href}%
\providecommand \@@href[1]{\endgroup#1\@@endlink}%
\providecommand \@sanitize@url [0]{\catcode `\\12\catcode `\$12\catcode `\&12\catcode `\#12\catcode `\^12\catcode `\_12\catcode `\%12\relax}%
\providecommand \@@startlink[1]{}%
\providecommand \@@endlink[0]{}%
\providecommand \url  [0]{\begingroup\@sanitize@url \@url }%
\providecommand \@url [1]{\endgroup\@href {#1}{\urlprefix }}%
\providecommand \urlprefix  [0]{URL }%
\providecommand \Eprint [0]{\href }%
\providecommand \doibase [0]{http://dx.doi.org/}%
\providecommand \selectlanguage [0]{\@gobble}%
\providecommand \bibinfo  [0]{\@secondoftwo}%
\providecommand \bibfield  [0]{\@secondoftwo}%
\providecommand \translation [1]{[#1]}%
\providecommand \BibitemOpen [0]{}%
\providecommand \bibitemStop [0]{}%
\providecommand \bibitemNoStop [0]{.\EOS\space}%
\providecommand \EOS [0]{\spacefactor3000\relax}%
\providecommand \BibitemShut  [1]{\csname bibitem#1\endcsname}%
\let\auto@bib@innerbib\@empty
\bibitem [{\citenamefont {Tajima}\ and\ \citenamefont {Dawson}(1979)}]{tajima79}%
  \BibitemOpen
  \bibfield  {author} {\bibinfo {author} {\bibfnamefont {T.}~\bibnamefont {Tajima}}\ and\ \bibinfo {author} {\bibfnamefont {J.~M.}\ \bibnamefont {Dawson}},\ }\href {\doibase 10.1103/PhysRevLett.43.267} {\bibfield  {journal} {\bibinfo  {journal} {Phys. Rev. Lett.}\ }\textbf {\bibinfo {volume} {43}},\ \bibinfo {pages} {267} (\bibinfo {year} {1979})}\BibitemShut {NoStop}%
\bibitem [{\citenamefont {Leemans}\ \emph {et~al.}(2006)\citenamefont {Leemans}, \citenamefont {Nagler}, \citenamefont {Gonsalves}, \citenamefont {T{\'o}th}, \citenamefont {Nakamura}, \citenamefont {Geddes}, \citenamefont {Esarey}, \citenamefont {Schroeder},\ and\ \citenamefont {Hooker}}]{leemans2006}%
  \BibitemOpen
  \bibfield  {author} {\bibinfo {author} {\bibfnamefont {W.~P.}\ \bibnamefont {Leemans}}, \bibinfo {author} {\bibfnamefont {B.}~\bibnamefont {Nagler}}, \bibinfo {author} {\bibfnamefont {A.~J.}\ \bibnamefont {Gonsalves}}, \bibinfo {author} {\bibfnamefont {C.}~\bibnamefont {T{\'o}th}}, \bibinfo {author} {\bibfnamefont {K.}~\bibnamefont {Nakamura}}, \bibinfo {author} {\bibfnamefont {C.~G.~R.}\ \bibnamefont {Geddes}}, \bibinfo {author} {\bibfnamefont {E.}~\bibnamefont {Esarey}}, \bibinfo {author} {\bibfnamefont {C.~B.}\ \bibnamefont {Schroeder}}, \ and\ \bibinfo {author} {\bibfnamefont {S.~M.}\ \bibnamefont {Hooker}},\ }\href {\doibase 10.1038/nphys418} {\bibfield  {journal} {\bibinfo  {journal} {Nature Physics}\ }\textbf {\bibinfo {volume} {2}},\ \bibinfo {pages} {696} (\bibinfo {year} {2006})}\BibitemShut {NoStop}%
\bibitem [{\citenamefont {Kneip}\ \emph {et~al.}(2009)\citenamefont {Kneip}, \citenamefont {Nagel}, \citenamefont {Martins}, \citenamefont {Mangles}, \citenamefont {Bellei}, \citenamefont {Chekhlov}, \citenamefont {Clarke}, \citenamefont {Delerue}, \citenamefont {Divall}, \citenamefont {Doucas}, \citenamefont {Ertel}, \citenamefont {Fiuza}, \citenamefont {Fonseca}, \citenamefont {Foster}, \citenamefont {Hawkes}, \citenamefont {Hooker}, \citenamefont {Krushelnick}, \citenamefont {Mori}, \citenamefont {Palmer}, \citenamefont {Phuoc}, \citenamefont {Rajeev}, \citenamefont {Schreiber}, \citenamefont {Streeter}, \citenamefont {Urner}, \citenamefont {Vieira}, \citenamefont {Silva},\ and\ \citenamefont {Najmudin}}]{Kneip.2009}%
  \BibitemOpen
  \bibfield  {author} {\bibinfo {author} {\bibfnamefont {S.}~\bibnamefont {Kneip}}, \bibinfo {author} {\bibfnamefont {S.}~\bibnamefont {Nagel}}, \bibinfo {author} {\bibfnamefont {S.}~\bibnamefont {Martins}}, \bibinfo {author} {\bibfnamefont {S.}~\bibnamefont {Mangles}}, \bibinfo {author} {\bibfnamefont {C.}~\bibnamefont {Bellei}}, \bibinfo {author} {\bibfnamefont {O.}~\bibnamefont {Chekhlov}}, \bibinfo {author} {\bibfnamefont {R.}~\bibnamefont {Clarke}}, \bibinfo {author} {\bibfnamefont {N.}~\bibnamefont {Delerue}}, \bibinfo {author} {\bibfnamefont {E.}~\bibnamefont {Divall}}, \bibinfo {author} {\bibfnamefont {G.}~\bibnamefont {Doucas}}, \bibinfo {author} {\bibfnamefont {K.}~\bibnamefont {Ertel}}, \bibinfo {author} {\bibfnamefont {F.}~\bibnamefont {Fiuza}}, \bibinfo {author} {\bibfnamefont {R.}~\bibnamefont {Fonseca}}, \bibinfo {author} {\bibfnamefont {P.}~\bibnamefont {Foster}}, \bibinfo {author} {\bibfnamefont {S.}~\bibnamefont {Hawkes}}, \bibinfo {author} {\bibfnamefont {C.}~\bibnamefont {Hooker}},
  \bibinfo {author} {\bibfnamefont {K.}~\bibnamefont {Krushelnick}}, \bibinfo {author} {\bibfnamefont {W.}~\bibnamefont {Mori}}, \bibinfo {author} {\bibfnamefont {C.}~\bibnamefont {Palmer}}, \bibinfo {author} {\bibfnamefont {K.}~\bibnamefont {Phuoc}}, \bibinfo {author} {\bibfnamefont {P.}~\bibnamefont {Rajeev}}, \bibinfo {author} {\bibfnamefont {J.}~\bibnamefont {Schreiber}}, \bibinfo {author} {\bibfnamefont {M.}~\bibnamefont {Streeter}}, \bibinfo {author} {\bibfnamefont {D.}~\bibnamefont {Urner}}, \bibinfo {author} {\bibfnamefont {J.}~\bibnamefont {Vieira}}, \bibinfo {author} {\bibfnamefont {L.}~\bibnamefont {Silva}}, \ and\ \bibinfo {author} {\bibfnamefont {Z.}~\bibnamefont {Najmudin}},\ }\href {\doibase 10.1103/physrevlett.103.035002} {\bibfield  {journal} {\bibinfo  {journal} {Physical Review Letters}\ }\textbf {\bibinfo {volume} {103}},\ \bibinfo {pages} {035002} (\bibinfo {year} {2009})}\BibitemShut {NoStop}%
\bibitem [{\citenamefont {Wang}\ \emph {et~al.}(2013)\citenamefont {Wang}, \citenamefont {Zgadzaj}, \citenamefont {Fazel}, \citenamefont {Li}, \citenamefont {Yi}, \citenamefont {Zhang}, \citenamefont {Henderson}, \citenamefont {Chang}, \citenamefont {Korzekwa}, \citenamefont {Tsai}, \citenamefont {Pai}, \citenamefont {Quevedo}, \citenamefont {Dyer}, \citenamefont {Gaul}, \citenamefont {Martinez}, \citenamefont {Bernstein}, \citenamefont {Borger}, \citenamefont {Spinks}, \citenamefont {Donovan}, \citenamefont {Khudik}, \citenamefont {Shvets}, \citenamefont {Ditmire},\ and\ \citenamefont {Downer}}]{Wang.2013hwz}%
  \BibitemOpen
  \bibfield  {author} {\bibinfo {author} {\bibfnamefont {X.}~\bibnamefont {Wang}}, \bibinfo {author} {\bibfnamefont {R.}~\bibnamefont {Zgadzaj}}, \bibinfo {author} {\bibfnamefont {N.}~\bibnamefont {Fazel}}, \bibinfo {author} {\bibfnamefont {Z.}~\bibnamefont {Li}}, \bibinfo {author} {\bibfnamefont {S.~A.}\ \bibnamefont {Yi}}, \bibinfo {author} {\bibfnamefont {X.}~\bibnamefont {Zhang}}, \bibinfo {author} {\bibfnamefont {W.}~\bibnamefont {Henderson}}, \bibinfo {author} {\bibfnamefont {Y.~Y.}\ \bibnamefont {Chang}}, \bibinfo {author} {\bibfnamefont {R.}~\bibnamefont {Korzekwa}}, \bibinfo {author} {\bibfnamefont {H.~E.}\ \bibnamefont {Tsai}}, \bibinfo {author} {\bibfnamefont {C.~H.}\ \bibnamefont {Pai}}, \bibinfo {author} {\bibfnamefont {H.}~\bibnamefont {Quevedo}}, \bibinfo {author} {\bibfnamefont {G.}~\bibnamefont {Dyer}}, \bibinfo {author} {\bibfnamefont {E.}~\bibnamefont {Gaul}}, \bibinfo {author} {\bibfnamefont {M.}~\bibnamefont {Martinez}}, \bibinfo {author} {\bibfnamefont {A.~C.}\ \bibnamefont {Bernstein}},
  \bibinfo {author} {\bibfnamefont {T.}~\bibnamefont {Borger}}, \bibinfo {author} {\bibfnamefont {M.}~\bibnamefont {Spinks}}, \bibinfo {author} {\bibfnamefont {M.}~\bibnamefont {Donovan}}, \bibinfo {author} {\bibfnamefont {V.}~\bibnamefont {Khudik}}, \bibinfo {author} {\bibfnamefont {G.}~\bibnamefont {Shvets}}, \bibinfo {author} {\bibfnamefont {T.}~\bibnamefont {Ditmire}}, \ and\ \bibinfo {author} {\bibfnamefont {M.~C.}\ \bibnamefont {Downer}},\ }\href {\doibase 10.1038/ncomms2988} {\bibfield  {journal} {\bibinfo  {journal} {Nature Communications}\ }\textbf {\bibinfo {volume} {4}} (\bibinfo {year} {2013}),\ 10.1038/ncomms2988}\BibitemShut {NoStop}%
\bibitem [{\citenamefont {Leemans}\ \emph {et~al.}(2014)\citenamefont {Leemans}, \citenamefont {Gonsalves}, \citenamefont {Mao}, \citenamefont {Nakamura}, \citenamefont {Benedetti}, \citenamefont {Schroeder}, \citenamefont {T\'oth}, \citenamefont {Daniels}, \citenamefont {Mittelberger}, \citenamefont {Bulanov}, \citenamefont {Vay}, \citenamefont {Geddes},\ and\ \citenamefont {Esarey}}]{leemans2014}%
  \BibitemOpen
  \bibfield  {author} {\bibinfo {author} {\bibfnamefont {W.~P.}\ \bibnamefont {Leemans}}, \bibinfo {author} {\bibfnamefont {A.~J.}\ \bibnamefont {Gonsalves}}, \bibinfo {author} {\bibfnamefont {H.-S.}\ \bibnamefont {Mao}}, \bibinfo {author} {\bibfnamefont {K.}~\bibnamefont {Nakamura}}, \bibinfo {author} {\bibfnamefont {C.}~\bibnamefont {Benedetti}}, \bibinfo {author} {\bibfnamefont {C.~B.}\ \bibnamefont {Schroeder}}, \bibinfo {author} {\bibfnamefont {C.}~\bibnamefont {T\'oth}}, \bibinfo {author} {\bibfnamefont {J.}~\bibnamefont {Daniels}}, \bibinfo {author} {\bibfnamefont {D.~E.}\ \bibnamefont {Mittelberger}}, \bibinfo {author} {\bibfnamefont {S.~S.}\ \bibnamefont {Bulanov}}, \bibinfo {author} {\bibfnamefont {J.-L.}\ \bibnamefont {Vay}}, \bibinfo {author} {\bibfnamefont {C.~G.~R.}\ \bibnamefont {Geddes}}, \ and\ \bibinfo {author} {\bibfnamefont {E.}~\bibnamefont {Esarey}},\ }\href {\doibase 10.1103/PhysRevLett.113.245002} {\bibfield  {journal} {\bibinfo  {journal} {Phys. Rev. Lett.}\ }\textbf {\bibinfo
  {volume} {113}},\ \bibinfo {pages} {245002} (\bibinfo {year} {2014})}\BibitemShut {NoStop}%
\bibitem [{\citenamefont {Shin}\ \emph {et~al.}(2018)\citenamefont {Shin}, \citenamefont {Kim}, \citenamefont {Pathak}, \citenamefont {Hojbota}, \citenamefont {Lee}, \citenamefont {Sung}, \citenamefont {Lee}, \citenamefont {Yoon}, \citenamefont {Jeon}, \citenamefont {Nakajima}, \citenamefont {Sylla}, \citenamefont {Lifschitz}, \citenamefont {Guillaume}, \citenamefont {Thaury}, \citenamefont {Malka},\ and\ \citenamefont {Nam}}]{Shin.2018}%
  \BibitemOpen
  \bibfield  {author} {\bibinfo {author} {\bibfnamefont {J.}~\bibnamefont {Shin}}, \bibinfo {author} {\bibfnamefont {H.~T.}\ \bibnamefont {Kim}}, \bibinfo {author} {\bibfnamefont {V.~B.}\ \bibnamefont {Pathak}}, \bibinfo {author} {\bibfnamefont {C.}~\bibnamefont {Hojbota}}, \bibinfo {author} {\bibfnamefont {S.~K.}\ \bibnamefont {Lee}}, \bibinfo {author} {\bibfnamefont {J.~H.}\ \bibnamefont {Sung}}, \bibinfo {author} {\bibfnamefont {H.~W.}\ \bibnamefont {Lee}}, \bibinfo {author} {\bibfnamefont {J.~W.}\ \bibnamefont {Yoon}}, \bibinfo {author} {\bibfnamefont {C.}~\bibnamefont {Jeon}}, \bibinfo {author} {\bibfnamefont {K.}~\bibnamefont {Nakajima}}, \bibinfo {author} {\bibfnamefont {F.}~\bibnamefont {Sylla}}, \bibinfo {author} {\bibfnamefont {A.}~\bibnamefont {Lifschitz}}, \bibinfo {author} {\bibfnamefont {E.}~\bibnamefont {Guillaume}}, \bibinfo {author} {\bibfnamefont {C.}~\bibnamefont {Thaury}}, \bibinfo {author} {\bibfnamefont {V.}~\bibnamefont {Malka}}, \ and\ \bibinfo {author} {\bibfnamefont {C.~H.}\
  \bibnamefont {Nam}},\ }\href {\doibase 10.1088/1361-6587/aabd10} {\bibfield  {journal} {\bibinfo  {journal} {Plasma Physics and Controlled Fusion}\ }\textbf {\bibinfo {volume} {60}},\ \bibinfo {pages} {064007 } (\bibinfo {year} {2018})}\BibitemShut {NoStop}%
\bibitem [{\citenamefont {Gonsalves}\ \emph {et~al.}(2019)\citenamefont {Gonsalves}, \citenamefont {Nakamura}, \citenamefont {Daniels}, \citenamefont {Benedetti}, \citenamefont {Pieronek}, \citenamefont {de~Raadt}, \citenamefont {Steinke}, \citenamefont {Bin}, \citenamefont {Bulanov}, \citenamefont {van Tilborg}, \citenamefont {Geddes}, \citenamefont {Schroeder}, \citenamefont {T\'oth}, \citenamefont {Esarey}, \citenamefont {Swanson}, \citenamefont {Fan-Chiang}, \citenamefont {Bagdasarov}, \citenamefont {Bobrova}, \citenamefont {Gasilov}, \citenamefont {Korn}, \citenamefont {Sasorov},\ and\ \citenamefont {Leemans}}]{gonsalves2019}%
  \BibitemOpen
  \bibfield  {author} {\bibinfo {author} {\bibfnamefont {A.~J.}\ \bibnamefont {Gonsalves}}, \bibinfo {author} {\bibfnamefont {K.}~\bibnamefont {Nakamura}}, \bibinfo {author} {\bibfnamefont {J.}~\bibnamefont {Daniels}}, \bibinfo {author} {\bibfnamefont {C.}~\bibnamefont {Benedetti}}, \bibinfo {author} {\bibfnamefont {C.}~\bibnamefont {Pieronek}}, \bibinfo {author} {\bibfnamefont {T.~C.~H.}\ \bibnamefont {de~Raadt}}, \bibinfo {author} {\bibfnamefont {S.}~\bibnamefont {Steinke}}, \bibinfo {author} {\bibfnamefont {J.~H.}\ \bibnamefont {Bin}}, \bibinfo {author} {\bibfnamefont {S.~S.}\ \bibnamefont {Bulanov}}, \bibinfo {author} {\bibfnamefont {J.}~\bibnamefont {van Tilborg}}, \bibinfo {author} {\bibfnamefont {C.~G.~R.}\ \bibnamefont {Geddes}}, \bibinfo {author} {\bibfnamefont {C.~B.}\ \bibnamefont {Schroeder}}, \bibinfo {author} {\bibfnamefont {C.}~\bibnamefont {T\'oth}}, \bibinfo {author} {\bibfnamefont {E.}~\bibnamefont {Esarey}}, \bibinfo {author} {\bibfnamefont {K.}~\bibnamefont {Swanson}}, \bibinfo {author}
  {\bibfnamefont {L.}~\bibnamefont {Fan-Chiang}}, \bibinfo {author} {\bibfnamefont {G.}~\bibnamefont {Bagdasarov}}, \bibinfo {author} {\bibfnamefont {N.}~\bibnamefont {Bobrova}}, \bibinfo {author} {\bibfnamefont {V.}~\bibnamefont {Gasilov}}, \bibinfo {author} {\bibfnamefont {G.}~\bibnamefont {Korn}}, \bibinfo {author} {\bibfnamefont {P.}~\bibnamefont {Sasorov}}, \ and\ \bibinfo {author} {\bibfnamefont {W.~P.}\ \bibnamefont {Leemans}},\ }\href {\doibase 10.1103/PhysRevLett.122.084801} {\bibfield  {journal} {\bibinfo  {journal} {Phys. Rev. Lett.}\ }\textbf {\bibinfo {volume} {122}},\ \bibinfo {pages} {084801} (\bibinfo {year} {2019})}\BibitemShut {NoStop}%
\bibitem [{\citenamefont {Miao}\ \emph {et~al.}(2022)\citenamefont {Miao}, \citenamefont {Shrock}, \citenamefont {Feder}, \citenamefont {Hollinger}, \citenamefont {Morrison}, \citenamefont {Nedbailo}, \citenamefont {Picksley}, \citenamefont {Song}, \citenamefont {Wang}, \citenamefont {Rocca},\ and\ \citenamefont {Milchberg}}]{miao2022}%
  \BibitemOpen
  \bibfield  {author} {\bibinfo {author} {\bibfnamefont {B.}~\bibnamefont {Miao}}, \bibinfo {author} {\bibfnamefont {J.~E.}\ \bibnamefont {Shrock}}, \bibinfo {author} {\bibfnamefont {L.}~\bibnamefont {Feder}}, \bibinfo {author} {\bibfnamefont {R.~C.}\ \bibnamefont {Hollinger}}, \bibinfo {author} {\bibfnamefont {J.}~\bibnamefont {Morrison}}, \bibinfo {author} {\bibfnamefont {R.}~\bibnamefont {Nedbailo}}, \bibinfo {author} {\bibfnamefont {A.}~\bibnamefont {Picksley}}, \bibinfo {author} {\bibfnamefont {H.}~\bibnamefont {Song}}, \bibinfo {author} {\bibfnamefont {S.}~\bibnamefont {Wang}}, \bibinfo {author} {\bibfnamefont {J.~J.}\ \bibnamefont {Rocca}}, \ and\ \bibinfo {author} {\bibfnamefont {H.~M.}\ \bibnamefont {Milchberg}},\ }\href {\doibase 10.1103/PhysRevX.12.031038} {\bibfield  {journal} {\bibinfo  {journal} {Phys. Rev. X}\ }\textbf {\bibinfo {volume} {12}},\ \bibinfo {pages} {031038} (\bibinfo {year} {2022})}\BibitemShut {NoStop}%
\bibitem [{\citenamefont {Oubrerie}\ \emph {et~al.}(2022)\citenamefont {Oubrerie}, \citenamefont {Leblanc}, \citenamefont {Kononenko}, \citenamefont {Lahaye}, \citenamefont {Andriyash}, \citenamefont {Gautier}, \citenamefont {Goddet}, \citenamefont {Martelli}, \citenamefont {Tafzi}, \citenamefont {Ta~Phuoc}, \citenamefont {Smartsev},\ and\ \citenamefont {Thaury}}]{Oubrerie2022LSA}%
  \BibitemOpen
  \bibfield  {author} {\bibinfo {author} {\bibfnamefont {K.}~\bibnamefont {Oubrerie}}, \bibinfo {author} {\bibfnamefont {A.}~\bibnamefont {Leblanc}}, \bibinfo {author} {\bibfnamefont {O.}~\bibnamefont {Kononenko}}, \bibinfo {author} {\bibfnamefont {R.}~\bibnamefont {Lahaye}}, \bibinfo {author} {\bibfnamefont {I.~A.}\ \bibnamefont {Andriyash}}, \bibinfo {author} {\bibfnamefont {J.}~\bibnamefont {Gautier}}, \bibinfo {author} {\bibfnamefont {J.-P.}\ \bibnamefont {Goddet}}, \bibinfo {author} {\bibfnamefont {L.}~\bibnamefont {Martelli}}, \bibinfo {author} {\bibfnamefont {A.}~\bibnamefont {Tafzi}}, \bibinfo {author} {\bibfnamefont {K.}~\bibnamefont {Ta~Phuoc}}, \bibinfo {author} {\bibfnamefont {S.}~\bibnamefont {Smartsev}}, \ and\ \bibinfo {author} {\bibfnamefont {C.}~\bibnamefont {Thaury}},\ }\href {\doibase 10.1038/s41377-022-00862-0} {\bibfield  {journal} {\bibinfo  {journal} {Light: Science {\&} Applications}\ }\textbf {\bibinfo {volume} {11}},\ \bibinfo {pages} {180} (\bibinfo {year} {2022})}\BibitemShut
  {NoStop}%
\bibitem [{\citenamefont {Picksley}\ \emph {et~al.}(2023)\citenamefont {Picksley}, \citenamefont {Chappell}, \citenamefont {Archer}, \citenamefont {Bourgeois}, \citenamefont {Cowley}, \citenamefont {Emerson}, \citenamefont {Feder}, \citenamefont {Gu}, \citenamefont {Jakobsson}, \citenamefont {Ross}, \citenamefont {Wang}, \citenamefont {Walczak},\ and\ \citenamefont {Hooker}}]{Picksley2023}%
  \BibitemOpen
  \bibfield  {author} {\bibinfo {author} {\bibfnamefont {A.}~\bibnamefont {Picksley}}, \bibinfo {author} {\bibfnamefont {J.}~\bibnamefont {Chappell}}, \bibinfo {author} {\bibfnamefont {E.}~\bibnamefont {Archer}}, \bibinfo {author} {\bibfnamefont {N.}~\bibnamefont {Bourgeois}}, \bibinfo {author} {\bibfnamefont {J.}~\bibnamefont {Cowley}}, \bibinfo {author} {\bibfnamefont {D.~R.}\ \bibnamefont {Emerson}}, \bibinfo {author} {\bibfnamefont {L.}~\bibnamefont {Feder}}, \bibinfo {author} {\bibfnamefont {X.~J.}\ \bibnamefont {Gu}}, \bibinfo {author} {\bibfnamefont {O.}~\bibnamefont {Jakobsson}}, \bibinfo {author} {\bibfnamefont {A.~J.}\ \bibnamefont {Ross}}, \bibinfo {author} {\bibfnamefont {W.}~\bibnamefont {Wang}}, \bibinfo {author} {\bibfnamefont {R.}~\bibnamefont {Walczak}}, \ and\ \bibinfo {author} {\bibfnamefont {S.~M.}\ \bibnamefont {Hooker}},\ }\href@noop {} {\enquote {\bibinfo {title} {{All-optical GeV electron bunch generation in a laser-plasma accelerator via truncated-channel injection}},}\ } (\bibinfo
  {year} {2023}),\ \Eprint {http://arxiv.org/abs/2307.13689} {arXiv:2307.13689 [physics.acc-ph]} \BibitemShut {NoStop}%
\bibitem [{\citenamefont {Corde}\ \emph {et~al.}(2013)\citenamefont {Corde}, \citenamefont {Phuoc}, \citenamefont {Lambert}, \citenamefont {Fitour}, \citenamefont {Malka}, \citenamefont {Rousse}, \citenamefont {Beck},\ and\ \citenamefont {Lefebvre}}]{Corde.2013}%
  \BibitemOpen
  \bibfield  {author} {\bibinfo {author} {\bibfnamefont {S.}~\bibnamefont {Corde}}, \bibinfo {author} {\bibfnamefont {K.~T.}\ \bibnamefont {Phuoc}}, \bibinfo {author} {\bibfnamefont {G.}~\bibnamefont {Lambert}}, \bibinfo {author} {\bibfnamefont {R.}~\bibnamefont {Fitour}}, \bibinfo {author} {\bibfnamefont {V.}~\bibnamefont {Malka}}, \bibinfo {author} {\bibfnamefont {A.}~\bibnamefont {Rousse}}, \bibinfo {author} {\bibfnamefont {A.}~\bibnamefont {Beck}}, \ and\ \bibinfo {author} {\bibfnamefont {E.}~\bibnamefont {Lefebvre}},\ }\href {\doibase 10.1103/revmodphys.85.1} {\bibfield  {journal} {\bibinfo  {journal} {Reviews of Modern Physics}\ }\textbf {\bibinfo {volume} {85}},\ \bibinfo {pages} {1 } (\bibinfo {year} {2013})}\BibitemShut {NoStop}%
\bibitem [{\citenamefont {Albert}\ and\ \citenamefont {Thomas}(2016)}]{Albert.2016}%
  \BibitemOpen
  \bibfield  {author} {\bibinfo {author} {\bibfnamefont {F.}~\bibnamefont {Albert}}\ and\ \bibinfo {author} {\bibfnamefont {A.~G.~R.}\ \bibnamefont {Thomas}},\ }\href {\doibase 10.1088/0741-3335/58/10/103001} {\bibfield  {journal} {\bibinfo  {journal} {Plasma Physics and Controlled Fusion}\ }\textbf {\bibinfo {volume} {58}},\ \bibinfo {pages} {103001} (\bibinfo {year} {2016})}\BibitemShut {NoStop}%
\bibitem [{\citenamefont {Wang}\ \emph {et~al.}(2021)\citenamefont {Wang}, \citenamefont {Feng}, \citenamefont {Ke}, \citenamefont {Yu}, \citenamefont {Xu}, \citenamefont {Qi}, \citenamefont {Chen}, \citenamefont {Qin}, \citenamefont {Zhang}, \citenamefont {Fang}, \citenamefont {Liu}, \citenamefont {Jiang}, \citenamefont {Wang}, \citenamefont {Wang}, \citenamefont {Yang}, \citenamefont {Wu}, \citenamefont {Leng}, \citenamefont {Liu}, \citenamefont {Li},\ and\ \citenamefont {Xu}}]{Wang.2021}%
  \BibitemOpen
  \bibfield  {author} {\bibinfo {author} {\bibfnamefont {W.}~\bibnamefont {Wang}}, \bibinfo {author} {\bibfnamefont {K.}~\bibnamefont {Feng}}, \bibinfo {author} {\bibfnamefont {L.}~\bibnamefont {Ke}}, \bibinfo {author} {\bibfnamefont {C.}~\bibnamefont {Yu}}, \bibinfo {author} {\bibfnamefont {Y.}~\bibnamefont {Xu}}, \bibinfo {author} {\bibfnamefont {R.}~\bibnamefont {Qi}}, \bibinfo {author} {\bibfnamefont {Y.}~\bibnamefont {Chen}}, \bibinfo {author} {\bibfnamefont {Z.}~\bibnamefont {Qin}}, \bibinfo {author} {\bibfnamefont {Z.}~\bibnamefont {Zhang}}, \bibinfo {author} {\bibfnamefont {M.}~\bibnamefont {Fang}}, \bibinfo {author} {\bibfnamefont {J.}~\bibnamefont {Liu}}, \bibinfo {author} {\bibfnamefont {K.}~\bibnamefont {Jiang}}, \bibinfo {author} {\bibfnamefont {H.}~\bibnamefont {Wang}}, \bibinfo {author} {\bibfnamefont {C.}~\bibnamefont {Wang}}, \bibinfo {author} {\bibfnamefont {X.}~\bibnamefont {Yang}}, \bibinfo {author} {\bibfnamefont {F.}~\bibnamefont {Wu}}, \bibinfo {author} {\bibfnamefont {Y.}~\bibnamefont
  {Leng}}, \bibinfo {author} {\bibfnamefont {J.}~\bibnamefont {Liu}}, \bibinfo {author} {\bibfnamefont {R.}~\bibnamefont {Li}}, \ and\ \bibinfo {author} {\bibfnamefont {Z.}~\bibnamefont {Xu}},\ }\href {\doibase 10.1038/s41586-021-03678-x} {\bibfield  {journal} {\bibinfo  {journal} {Nature}\ }\textbf {\bibinfo {volume} {595}},\ \bibinfo {pages} {516} (\bibinfo {year} {2021})}\BibitemShut {NoStop}%
\bibitem [{\citenamefont {Labat}\ \emph {et~al.}(2022)\citenamefont {Labat}, \citenamefont {Cabadağ}, \citenamefont {Ghaith}, \citenamefont {Irman}, \citenamefont {Berlioux}, \citenamefont {Berteaud}, \citenamefont {Blache}, \citenamefont {Bock}, \citenamefont {Bouvet}, \citenamefont {Briquez}, \citenamefont {Chang}, \citenamefont {Corde}, \citenamefont {Debus}, \citenamefont {Oliveira}, \citenamefont {Duval}, \citenamefont {Dietrich}, \citenamefont {Ajjouri}, \citenamefont {Eisenmann}, \citenamefont {Gautier}, \citenamefont {Gebhardt}, \citenamefont {Grams}, \citenamefont {Helbig}, \citenamefont {Herbeaux}, \citenamefont {Hubert}, \citenamefont {Kitegi}, \citenamefont {Kononenko}, \citenamefont {Kuntzsch}, \citenamefont {LaBerge}, \citenamefont {Lê}, \citenamefont {Leluan}, \citenamefont {Loulergue}, \citenamefont {Malka}, \citenamefont {Marteau}, \citenamefont {Guyen}, \citenamefont {Oumbarek-Espinos}, \citenamefont {Pausch}, \citenamefont {Pereira}, \citenamefont {Püschel}, \citenamefont {Ricaud},
  \citenamefont {Rommeluere}, \citenamefont {Roussel}, \citenamefont {Rousseau}, \citenamefont {Schöbel}, \citenamefont {Sebdaoui}, \citenamefont {Steiniger}, \citenamefont {Tavakoli}, \citenamefont {Thaury}, \citenamefont {Ufer}, \citenamefont {Valléau}, \citenamefont {Vandenberghe}, \citenamefont {Vétéran}, \citenamefont {Schramm},\ and\ \citenamefont {Couprie}}]{Labat.2022nd8}%
  \BibitemOpen
  \bibfield  {author} {\bibinfo {author} {\bibfnamefont {M.}~\bibnamefont {Labat}}, \bibinfo {author} {\bibfnamefont {J.~C.}\ \bibnamefont {Cabadağ}}, \bibinfo {author} {\bibfnamefont {A.}~\bibnamefont {Ghaith}}, \bibinfo {author} {\bibfnamefont {A.}~\bibnamefont {Irman}}, \bibinfo {author} {\bibfnamefont {A.}~\bibnamefont {Berlioux}}, \bibinfo {author} {\bibfnamefont {P.}~\bibnamefont {Berteaud}}, \bibinfo {author} {\bibfnamefont {F.}~\bibnamefont {Blache}}, \bibinfo {author} {\bibfnamefont {S.}~\bibnamefont {Bock}}, \bibinfo {author} {\bibfnamefont {F.}~\bibnamefont {Bouvet}}, \bibinfo {author} {\bibfnamefont {F.}~\bibnamefont {Briquez}}, \bibinfo {author} {\bibfnamefont {Y.-Y.}\ \bibnamefont {Chang}}, \bibinfo {author} {\bibfnamefont {S.}~\bibnamefont {Corde}}, \bibinfo {author} {\bibfnamefont {A.}~\bibnamefont {Debus}}, \bibinfo {author} {\bibfnamefont {C.~D.}\ \bibnamefont {Oliveira}}, \bibinfo {author} {\bibfnamefont {J.-P.}\ \bibnamefont {Duval}}, \bibinfo {author} {\bibfnamefont {Y.}~\bibnamefont
  {Dietrich}}, \bibinfo {author} {\bibfnamefont {M.~E.}\ \bibnamefont {Ajjouri}}, \bibinfo {author} {\bibfnamefont {C.}~\bibnamefont {Eisenmann}}, \bibinfo {author} {\bibfnamefont {J.}~\bibnamefont {Gautier}}, \bibinfo {author} {\bibfnamefont {R.}~\bibnamefont {Gebhardt}}, \bibinfo {author} {\bibfnamefont {S.}~\bibnamefont {Grams}}, \bibinfo {author} {\bibfnamefont {U.}~\bibnamefont {Helbig}}, \bibinfo {author} {\bibfnamefont {C.}~\bibnamefont {Herbeaux}}, \bibinfo {author} {\bibfnamefont {N.}~\bibnamefont {Hubert}}, \bibinfo {author} {\bibfnamefont {C.}~\bibnamefont {Kitegi}}, \bibinfo {author} {\bibfnamefont {O.}~\bibnamefont {Kononenko}}, \bibinfo {author} {\bibfnamefont {M.}~\bibnamefont {Kuntzsch}}, \bibinfo {author} {\bibfnamefont {M.}~\bibnamefont {LaBerge}}, \bibinfo {author} {\bibfnamefont {S.}~\bibnamefont {Lê}}, \bibinfo {author} {\bibfnamefont {B.}~\bibnamefont {Leluan}}, \bibinfo {author} {\bibfnamefont {A.}~\bibnamefont {Loulergue}}, \bibinfo {author} {\bibfnamefont {V.}~\bibnamefont {Malka}},
  \bibinfo {author} {\bibfnamefont {F.}~\bibnamefont {Marteau}}, \bibinfo {author} {\bibfnamefont {M.~H.~N.}\ \bibnamefont {Guyen}}, \bibinfo {author} {\bibfnamefont {D.}~\bibnamefont {Oumbarek-Espinos}}, \bibinfo {author} {\bibfnamefont {R.}~\bibnamefont {Pausch}}, \bibinfo {author} {\bibfnamefont {D.}~\bibnamefont {Pereira}}, \bibinfo {author} {\bibfnamefont {T.}~\bibnamefont {Püschel}}, \bibinfo {author} {\bibfnamefont {J.-P.}\ \bibnamefont {Ricaud}}, \bibinfo {author} {\bibfnamefont {P.}~\bibnamefont {Rommeluere}}, \bibinfo {author} {\bibfnamefont {E.}~\bibnamefont {Roussel}}, \bibinfo {author} {\bibfnamefont {P.}~\bibnamefont {Rousseau}}, \bibinfo {author} {\bibfnamefont {S.}~\bibnamefont {Schöbel}}, \bibinfo {author} {\bibfnamefont {M.}~\bibnamefont {Sebdaoui}}, \bibinfo {author} {\bibfnamefont {K.}~\bibnamefont {Steiniger}}, \bibinfo {author} {\bibfnamefont {K.}~\bibnamefont {Tavakoli}}, \bibinfo {author} {\bibfnamefont {C.}~\bibnamefont {Thaury}}, \bibinfo {author} {\bibfnamefont {P.}~\bibnamefont
  {Ufer}}, \bibinfo {author} {\bibfnamefont {M.}~\bibnamefont {Valléau}}, \bibinfo {author} {\bibfnamefont {M.}~\bibnamefont {Vandenberghe}}, \bibinfo {author} {\bibfnamefont {J.}~\bibnamefont {Vétéran}}, \bibinfo {author} {\bibfnamefont {U.}~\bibnamefont {Schramm}}, \ and\ \bibinfo {author} {\bibfnamefont {M.-E.}\ \bibnamefont {Couprie}},\ }\href {\doibase 10.1038/s41566-022-01104-w} {\bibfield  {journal} {\bibinfo  {journal} {Nature Photonics}\ ,\ \bibinfo {pages} {1}} (\bibinfo {year} {2022})}\BibitemShut {NoStop}%
\bibitem [{\citenamefont {Strickland}\ and\ \citenamefont {Mourou}(1985)}]{strickland1985}%
  \BibitemOpen
  \bibfield  {author} {\bibinfo {author} {\bibfnamefont {D.}~\bibnamefont {Strickland}}\ and\ \bibinfo {author} {\bibfnamefont {G.}~\bibnamefont {Mourou}},\ }\href {\doibase https://doi.org/10.1016/0030-4018(85)90120-8} {\bibfield  {journal} {\bibinfo  {journal} {Optics Communications}\ }\textbf {\bibinfo {volume} {56}},\ \bibinfo {pages} {219} (\bibinfo {year} {1985})}\BibitemShut {NoStop}%
\bibitem [{\citenamefont {Siders}\ \emph {et~al.}(2019)\citenamefont {Siders}, \citenamefont {Galvin}, \citenamefont {Erlandson}, \citenamefont {Bayramian}, \citenamefont {Reagan}, \citenamefont {Sistrunk}, \citenamefont {Spinka},\ and\ \citenamefont {Haefner}}]{Siders.2019}%
  \BibitemOpen
  \bibfield  {author} {\bibinfo {author} {\bibnamefont {Siders}}, \bibinfo {author} {\bibnamefont {Galvin}}, \bibinfo {author} {\bibnamefont {Erlandson}}, \bibinfo {author} {\bibnamefont {Bayramian}}, \bibinfo {author} {\bibnamefont {Reagan}}, \bibinfo {author} {\bibnamefont {Sistrunk}}, \bibinfo {author} {\bibnamefont {Spinka}}, \ and\ \bibinfo {author} {\bibnamefont {Haefner}},\ }\href {\doibase 10.3390/instruments3030044} {\bibfield  {journal} {\bibinfo  {journal} {Instruments}\ }\textbf {\bibinfo {volume} {3}},\ \bibinfo {pages} {44 } (\bibinfo {year} {2019})}\BibitemShut {NoStop}%
\bibitem [{\citenamefont {Joshi}\ \emph {et~al.}(1984)\citenamefont {Joshi}, \citenamefont {Mori}, \citenamefont {Katsouleas}, \citenamefont {Dawson}, \citenamefont {Kindel},\ and\ \citenamefont {Forslund}}]{Joshi.1984}%
  \BibitemOpen
  \bibfield  {author} {\bibinfo {author} {\bibfnamefont {C.}~\bibnamefont {Joshi}}, \bibinfo {author} {\bibfnamefont {W.~B.}\ \bibnamefont {Mori}}, \bibinfo {author} {\bibfnamefont {T.}~\bibnamefont {Katsouleas}}, \bibinfo {author} {\bibfnamefont {J.~M.}\ \bibnamefont {Dawson}}, \bibinfo {author} {\bibfnamefont {J.~M.}\ \bibnamefont {Kindel}}, \ and\ \bibinfo {author} {\bibfnamefont {D.~W.}\ \bibnamefont {Forslund}},\ }\href {\doibase 10.1038/311525a0} {\bibfield  {journal} {\bibinfo  {journal} {Nature}\ }\textbf {\bibinfo {volume} {311}},\ \bibinfo {pages} {525 } (\bibinfo {year} {1984})}\BibitemShut {NoStop}%
\bibitem [{\citenamefont {Clayton}\ \emph {et~al.}(1993)\citenamefont {Clayton}, \citenamefont {Marsh}, \citenamefont {Dyson}, \citenamefont {Everett}, \citenamefont {Lal}, \citenamefont {Leemans}, \citenamefont {Williams},\ and\ \citenamefont {Joshi}}]{clayton1993}%
  \BibitemOpen
  \bibfield  {author} {\bibinfo {author} {\bibfnamefont {C.~E.}\ \bibnamefont {Clayton}}, \bibinfo {author} {\bibfnamefont {K.~A.}\ \bibnamefont {Marsh}}, \bibinfo {author} {\bibfnamefont {A.}~\bibnamefont {Dyson}}, \bibinfo {author} {\bibfnamefont {M.}~\bibnamefont {Everett}}, \bibinfo {author} {\bibfnamefont {A.}~\bibnamefont {Lal}}, \bibinfo {author} {\bibfnamefont {W.~P.}\ \bibnamefont {Leemans}}, \bibinfo {author} {\bibfnamefont {R.}~\bibnamefont {Williams}}, \ and\ \bibinfo {author} {\bibfnamefont {C.}~\bibnamefont {Joshi}},\ }\href {\doibase 10.1103/PhysRevLett.70.37} {\bibfield  {journal} {\bibinfo  {journal} {Phys. Rev. Lett.}\ }\textbf {\bibinfo {volume} {70}},\ \bibinfo {pages} {37} (\bibinfo {year} {1993})}\BibitemShut {NoStop}%
\bibitem [{\citenamefont {Tochitsky}\ \emph {et~al.}(2004)\citenamefont {Tochitsky}, \citenamefont {Narang}, \citenamefont {Filip}, \citenamefont {Musumeci}, \citenamefont {Clayton}, \citenamefont {Yoder}, \citenamefont {Marsh}, \citenamefont {Rosenzweig}, \citenamefont {Pellegrini},\ and\ \citenamefont {Joshi}}]{tochitsky2004}%
  \BibitemOpen
  \bibfield  {author} {\bibinfo {author} {\bibfnamefont {S.~Y.}\ \bibnamefont {Tochitsky}}, \bibinfo {author} {\bibfnamefont {R.}~\bibnamefont {Narang}}, \bibinfo {author} {\bibfnamefont {C.~V.}\ \bibnamefont {Filip}}, \bibinfo {author} {\bibfnamefont {P.}~\bibnamefont {Musumeci}}, \bibinfo {author} {\bibfnamefont {C.~E.}\ \bibnamefont {Clayton}}, \bibinfo {author} {\bibfnamefont {R.~B.}\ \bibnamefont {Yoder}}, \bibinfo {author} {\bibfnamefont {K.~A.}\ \bibnamefont {Marsh}}, \bibinfo {author} {\bibfnamefont {J.~B.}\ \bibnamefont {Rosenzweig}}, \bibinfo {author} {\bibfnamefont {C.}~\bibnamefont {Pellegrini}}, \ and\ \bibinfo {author} {\bibfnamefont {C.}~\bibnamefont {Joshi}},\ }\href {\doibase 10.1103/PhysRevLett.92.095004} {\bibfield  {journal} {\bibinfo  {journal} {Phys. Rev. Lett.}\ }\textbf {\bibinfo {volume} {92}},\ \bibinfo {pages} {095004} (\bibinfo {year} {2004})}\BibitemShut {NoStop}%
\bibitem [{Sup()}]{Supp_mat}%
  \BibitemOpen
  \href@noop {} {}\bibinfo {note} {{See Supplemental Material at \url{link-to-be-inserted} for further details on the experimental set-up, pulse train diagnosis and guiding measurements which includes Refs.~\cite{fournier2011,code_saturne,gmmBook,AIC,vandewetering2023stability,ADK}.}}\BibitemShut {Stop}%
\bibitem [{\citenamefont {Hooker}\ \emph {et~al.}(2014)\citenamefont {Hooker}, \citenamefont {Bartolini}, \citenamefont {Mangles}, \citenamefont {Tünnermann}, \citenamefont {Corner}, \citenamefont {Limpert}, \citenamefont {Seryi},\ and\ \citenamefont {Walczak}}]{Hooker_2014}%
  \BibitemOpen
  \bibfield  {author} {\bibinfo {author} {\bibfnamefont {S.~M.}\ \bibnamefont {Hooker}}, \bibinfo {author} {\bibfnamefont {R.}~\bibnamefont {Bartolini}}, \bibinfo {author} {\bibfnamefont {S.~P.~D.}\ \bibnamefont {Mangles}}, \bibinfo {author} {\bibfnamefont {A.}~\bibnamefont {Tünnermann}}, \bibinfo {author} {\bibfnamefont {L.}~\bibnamefont {Corner}}, \bibinfo {author} {\bibfnamefont {J.}~\bibnamefont {Limpert}}, \bibinfo {author} {\bibfnamefont {A.}~\bibnamefont {Seryi}}, \ and\ \bibinfo {author} {\bibfnamefont {R.}~\bibnamefont {Walczak}},\ }\href {\doibase 10.1088/0953-4075/47/23/234003} {\bibfield  {journal} {\bibinfo  {journal} {Journal of Physics B: Atomic, Molecular and Optical Physics}\ }\textbf {\bibinfo {volume} {47}},\ \bibinfo {pages} {234003} (\bibinfo {year} {2014})}\BibitemShut {NoStop}%
\bibitem [{\citenamefont {Wang}\ \emph {et~al.}(2020)\citenamefont {Wang}, \citenamefont {Chi}, \citenamefont {Baumgarten}, \citenamefont {Dehne}, \citenamefont {Meadows}, \citenamefont {Davenport}, \citenamefont {Murray}, \citenamefont {Reagan}, \citenamefont {Menoni},\ and\ \citenamefont {Rocca}}]{Wang.2020}%
  \BibitemOpen
  \bibfield  {author} {\bibinfo {author} {\bibfnamefont {Y.}~\bibnamefont {Wang}}, \bibinfo {author} {\bibfnamefont {H.}~\bibnamefont {Chi}}, \bibinfo {author} {\bibfnamefont {C.}~\bibnamefont {Baumgarten}}, \bibinfo {author} {\bibfnamefont {K.}~\bibnamefont {Dehne}}, \bibinfo {author} {\bibfnamefont {A.~R.}\ \bibnamefont {Meadows}}, \bibinfo {author} {\bibfnamefont {A.}~\bibnamefont {Davenport}}, \bibinfo {author} {\bibfnamefont {G.}~\bibnamefont {Murray}}, \bibinfo {author} {\bibfnamefont {B.~A.}\ \bibnamefont {Reagan}}, \bibinfo {author} {\bibfnamefont {C.~S.}\ \bibnamefont {Menoni}}, \ and\ \bibinfo {author} {\bibfnamefont {J.~J.}\ \bibnamefont {Rocca}},\ }\href {\doibase 10.1364/ol.413129} {\bibfield  {journal} {\bibinfo  {journal} {Optics Letters}\ }\textbf {\bibinfo {volume} {45}},\ \bibinfo {pages} {6615 } (\bibinfo {year} {2020})}\BibitemShut {NoStop}%
\bibitem [{\citenamefont {Jakobsson}\ \emph {et~al.}(2021)\citenamefont {Jakobsson}, \citenamefont {Hooker},\ and\ \citenamefont {Walczak}}]{jakobssonPMoPA}%
  \BibitemOpen
  \bibfield  {author} {\bibinfo {author} {\bibfnamefont {O.}~\bibnamefont {Jakobsson}}, \bibinfo {author} {\bibfnamefont {S.~M.}\ \bibnamefont {Hooker}}, \ and\ \bibinfo {author} {\bibfnamefont {R.}~\bibnamefont {Walczak}},\ }\href {\doibase 10.1103/PhysRevLett.127.184801} {\bibfield  {journal} {\bibinfo  {journal} {Phys. Rev. Lett.}\ }\textbf {\bibinfo {volume} {127}},\ \bibinfo {pages} {184801} (\bibinfo {year} {2021})}\BibitemShut {NoStop}%
\bibitem [{\citenamefont {Shalloo}\ \emph {et~al.}(2016)\citenamefont {Shalloo}, \citenamefont {Corner}, \citenamefont {Arran}, \citenamefont {Cowley}, \citenamefont {Cheung}, \citenamefont {Thornton}, \citenamefont {Walczak},\ and\ \citenamefont {Hooker}}]{Shalloo2016}%
  \BibitemOpen
  \bibfield  {author} {\bibinfo {author} {\bibfnamefont {R.}~\bibnamefont {Shalloo}}, \bibinfo {author} {\bibfnamefont {L.}~\bibnamefont {Corner}}, \bibinfo {author} {\bibfnamefont {C.}~\bibnamefont {Arran}}, \bibinfo {author} {\bibfnamefont {J.}~\bibnamefont {Cowley}}, \bibinfo {author} {\bibfnamefont {G.}~\bibnamefont {Cheung}}, \bibinfo {author} {\bibfnamefont {C.}~\bibnamefont {Thornton}}, \bibinfo {author} {\bibfnamefont {R.}~\bibnamefont {Walczak}}, \ and\ \bibinfo {author} {\bibfnamefont {S.}~\bibnamefont {Hooker}},\ }\href {\doibase https://doi.org/10.1016/j.nima.2016.02.044} {\bibfield  {journal} {\bibinfo  {journal} {Nuclear Instruments and Methods in Physics Research Section A: Accelerators, Spectrometers, Detectors and Associated Equipment}\ }\textbf {\bibinfo {volume} {829}},\ \bibinfo {pages} {383} (\bibinfo {year} {2016})},\ \bibinfo {note} {2nd European Advanced Accelerator Concepts Workshop - EAAC 2015}\BibitemShut {NoStop}%
\bibitem [{\citenamefont {Cowley}\ \emph {et~al.}(2017)\citenamefont {Cowley}, \citenamefont {Thornton}, \citenamefont {Arran}, \citenamefont {Shalloo}, \citenamefont {Corner}, \citenamefont {Cheung}, \citenamefont {Gregory}, \citenamefont {Mangles}, \citenamefont {Matlis}, \citenamefont {Symes}, \citenamefont {Walczak},\ and\ \citenamefont {Hooker}}]{cowley2017}%
  \BibitemOpen
  \bibfield  {author} {\bibinfo {author} {\bibfnamefont {J.}~\bibnamefont {Cowley}}, \bibinfo {author} {\bibfnamefont {C.}~\bibnamefont {Thornton}}, \bibinfo {author} {\bibfnamefont {C.}~\bibnamefont {Arran}}, \bibinfo {author} {\bibfnamefont {R.~J.}\ \bibnamefont {Shalloo}}, \bibinfo {author} {\bibfnamefont {L.}~\bibnamefont {Corner}}, \bibinfo {author} {\bibfnamefont {G.}~\bibnamefont {Cheung}}, \bibinfo {author} {\bibfnamefont {C.~D.}\ \bibnamefont {Gregory}}, \bibinfo {author} {\bibfnamefont {S.~P.~D.}\ \bibnamefont {Mangles}}, \bibinfo {author} {\bibfnamefont {N.~H.}\ \bibnamefont {Matlis}}, \bibinfo {author} {\bibfnamefont {D.~R.}\ \bibnamefont {Symes}}, \bibinfo {author} {\bibfnamefont {R.}~\bibnamefont {Walczak}}, \ and\ \bibinfo {author} {\bibfnamefont {S.~M.}\ \bibnamefont {Hooker}},\ }\href {\doibase 10.1103/PhysRevLett.119.044802} {\bibfield  {journal} {\bibinfo  {journal} {Phys. Rev. Lett.}\ }\textbf {\bibinfo {volume} {119}},\ \bibinfo {pages} {044802} (\bibinfo {year} {2017})}\BibitemShut
  {NoStop}%
\bibitem [{\citenamefont {Aniculaesei}\ \emph {et~al.}(2018)\citenamefont {Aniculaesei}, \citenamefont {Kim}, \citenamefont {Yoo}, \citenamefont {Oh},\ and\ \citenamefont {Nam}}]{Aniculaesei2018}%
  \BibitemOpen
  \bibfield  {author} {\bibinfo {author} {\bibfnamefont {C.}~\bibnamefont {Aniculaesei}}, \bibinfo {author} {\bibfnamefont {H.~T.}\ \bibnamefont {Kim}}, \bibinfo {author} {\bibfnamefont {B.~J.}\ \bibnamefont {Yoo}}, \bibinfo {author} {\bibfnamefont {K.~H.}\ \bibnamefont {Oh}}, \ and\ \bibinfo {author} {\bibfnamefont {C.~H.}\ \bibnamefont {Nam}},\ }\href {\doibase 10.1063/1.4993269} {\bibfield  {journal} {\bibinfo  {journal} {Review of Scientific Instruments}\ }\textbf {\bibinfo {volume} {89}},\ \bibinfo {pages} {025110} (\bibinfo {year} {2018})}\BibitemShut {NoStop}%
\bibitem [{\citenamefont {Shalloo}\ \emph {et~al.}(2018)\citenamefont {Shalloo}, \citenamefont {Arran}, \citenamefont {Corner}, \citenamefont {Holloway}, \citenamefont {Jonnerby}, \citenamefont {Walczak}, \citenamefont {Milchberg},\ and\ \citenamefont {Hooker}}]{shalloo2018}%
  \BibitemOpen
  \bibfield  {author} {\bibinfo {author} {\bibfnamefont {R.~J.}\ \bibnamefont {Shalloo}}, \bibinfo {author} {\bibfnamefont {C.}~\bibnamefont {Arran}}, \bibinfo {author} {\bibfnamefont {L.}~\bibnamefont {Corner}}, \bibinfo {author} {\bibfnamefont {J.}~\bibnamefont {Holloway}}, \bibinfo {author} {\bibfnamefont {J.}~\bibnamefont {Jonnerby}}, \bibinfo {author} {\bibfnamefont {R.}~\bibnamefont {Walczak}}, \bibinfo {author} {\bibfnamefont {H.~M.}\ \bibnamefont {Milchberg}}, \ and\ \bibinfo {author} {\bibfnamefont {S.~M.}\ \bibnamefont {Hooker}},\ }\href {\doibase 10.1103/PhysRevE.97.053203} {\bibfield  {journal} {\bibinfo  {journal} {Phys. Rev. E}\ }\textbf {\bibinfo {volume} {97}},\ \bibinfo {pages} {053203} (\bibinfo {year} {2018})}\BibitemShut {NoStop}%
\bibitem [{\citenamefont {Shalloo}\ \emph {et~al.}(2019)\citenamefont {Shalloo}, \citenamefont {Arran}, \citenamefont {Picksley}, \citenamefont {von Boetticher}, \citenamefont {Corner}, \citenamefont {Holloway}, \citenamefont {Hine}, \citenamefont {Jonnerby}, \citenamefont {Milchberg}, \citenamefont {Thornton}, \citenamefont {Walczak},\ and\ \citenamefont {Hooker}}]{shalloo2019}%
  \BibitemOpen
  \bibfield  {author} {\bibinfo {author} {\bibfnamefont {R.~J.}\ \bibnamefont {Shalloo}}, \bibinfo {author} {\bibfnamefont {C.}~\bibnamefont {Arran}}, \bibinfo {author} {\bibfnamefont {A.}~\bibnamefont {Picksley}}, \bibinfo {author} {\bibfnamefont {A.}~\bibnamefont {von Boetticher}}, \bibinfo {author} {\bibfnamefont {L.}~\bibnamefont {Corner}}, \bibinfo {author} {\bibfnamefont {J.}~\bibnamefont {Holloway}}, \bibinfo {author} {\bibfnamefont {G.}~\bibnamefont {Hine}}, \bibinfo {author} {\bibfnamefont {J.}~\bibnamefont {Jonnerby}}, \bibinfo {author} {\bibfnamefont {H.~M.}\ \bibnamefont {Milchberg}}, \bibinfo {author} {\bibfnamefont {C.}~\bibnamefont {Thornton}}, \bibinfo {author} {\bibfnamefont {R.}~\bibnamefont {Walczak}}, \ and\ \bibinfo {author} {\bibfnamefont {S.~M.}\ \bibnamefont {Hooker}},\ }\href {\doibase 10.1103/PhysRevAccelBeams.22.041302} {\bibfield  {journal} {\bibinfo  {journal} {Phys. Rev. Accel. Beams}\ }\textbf {\bibinfo {volume} {22}},\ \bibinfo {pages} {041302} (\bibinfo {year}
  {2019})}\BibitemShut {NoStop}%
\bibitem [{\citenamefont {Esarey}\ \emph {et~al.}(1990)\citenamefont {Esarey}, \citenamefont {Ting},\ and\ \citenamefont {Sprangle}}]{EsareyFreqShifts1990}%
  \BibitemOpen
  \bibfield  {author} {\bibinfo {author} {\bibfnamefont {E.}~\bibnamefont {Esarey}}, \bibinfo {author} {\bibfnamefont {A.}~\bibnamefont {Ting}}, \ and\ \bibinfo {author} {\bibfnamefont {P.}~\bibnamefont {Sprangle}},\ }\href {\doibase 10.1103/PhysRevA.42.3526} {\bibfield  {journal} {\bibinfo  {journal} {Phys. Rev. A}\ }\textbf {\bibinfo {volume} {42}},\ \bibinfo {pages} {3526} (\bibinfo {year} {1990})}\BibitemShut {NoStop}%
\bibitem [{\citenamefont {Picksley}\ \emph {et~al.}(2020{\natexlab{a}})\citenamefont {Picksley}, \citenamefont {Alejo}, \citenamefont {Shalloo}, \citenamefont {Arran}, \citenamefont {von Boetticher}, \citenamefont {Corner}, \citenamefont {Holloway}, \citenamefont {Jonnerby}, \citenamefont {Jakobsson}, \citenamefont {Thornton}, \citenamefont {Walczak},\ and\ \citenamefont {Hooker}}]{chofi2020}%
  \BibitemOpen
  \bibfield  {author} {\bibinfo {author} {\bibfnamefont {A.}~\bibnamefont {Picksley}}, \bibinfo {author} {\bibfnamefont {A.}~\bibnamefont {Alejo}}, \bibinfo {author} {\bibfnamefont {R.~J.}\ \bibnamefont {Shalloo}}, \bibinfo {author} {\bibfnamefont {C.}~\bibnamefont {Arran}}, \bibinfo {author} {\bibfnamefont {A.}~\bibnamefont {von Boetticher}}, \bibinfo {author} {\bibfnamefont {L.}~\bibnamefont {Corner}}, \bibinfo {author} {\bibfnamefont {J.~A.}\ \bibnamefont {Holloway}}, \bibinfo {author} {\bibfnamefont {J.}~\bibnamefont {Jonnerby}}, \bibinfo {author} {\bibfnamefont {O.}~\bibnamefont {Jakobsson}}, \bibinfo {author} {\bibfnamefont {C.}~\bibnamefont {Thornton}}, \bibinfo {author} {\bibfnamefont {R.}~\bibnamefont {Walczak}}, \ and\ \bibinfo {author} {\bibfnamefont {S.~M.}\ \bibnamefont {Hooker}},\ }\href {\doibase 10.1103/PhysRevE.102.053201} {\bibfield  {journal} {\bibinfo  {journal} {Phys. Rev. E}\ }\textbf {\bibinfo {volume} {102}},\ \bibinfo {pages} {053201} (\bibinfo {year}
  {2020}{\natexlab{a}})}\BibitemShut {NoStop}%
\bibitem [{\citenamefont {Feder}\ \emph {et~al.}(2020)\citenamefont {Feder}, \citenamefont {Miao}, \citenamefont {Shrock}, \citenamefont {Goffin},\ and\ \citenamefont {Milchberg}}]{feder2020}%
  \BibitemOpen
  \bibfield  {author} {\bibinfo {author} {\bibfnamefont {L.}~\bibnamefont {Feder}}, \bibinfo {author} {\bibfnamefont {B.}~\bibnamefont {Miao}}, \bibinfo {author} {\bibfnamefont {J.~E.}\ \bibnamefont {Shrock}}, \bibinfo {author} {\bibfnamefont {A.}~\bibnamefont {Goffin}}, \ and\ \bibinfo {author} {\bibfnamefont {H.~M.}\ \bibnamefont {Milchberg}},\ }\href {\doibase 10.1103/PhysRevResearch.2.043173} {\bibfield  {journal} {\bibinfo  {journal} {Phys. Rev. Res.}\ }\textbf {\bibinfo {volume} {2}},\ \bibinfo {pages} {043173} (\bibinfo {year} {2020})}\BibitemShut {NoStop}%
\bibitem [{\citenamefont {Max}\ \emph {et~al.}(1974)\citenamefont {Max}, \citenamefont {Arons},\ and\ \citenamefont {Langdon}}]{PhysRevLett.33.209}%
  \BibitemOpen
  \bibfield  {author} {\bibinfo {author} {\bibfnamefont {C.~E.}\ \bibnamefont {Max}}, \bibinfo {author} {\bibfnamefont {J.}~\bibnamefont {Arons}}, \ and\ \bibinfo {author} {\bibfnamefont {A.~B.}\ \bibnamefont {Langdon}},\ }\href {\doibase 10.1103/PhysRevLett.33.209} {\bibfield  {journal} {\bibinfo  {journal} {Phys. Rev. Lett.}\ }\textbf {\bibinfo {volume} {33}},\ \bibinfo {pages} {209} (\bibinfo {year} {1974})}\BibitemShut {NoStop}%
\bibitem [{\citenamefont {Esarey}\ \emph {et~al.}(1994)\citenamefont {Esarey}, \citenamefont {Krall},\ and\ \citenamefont {Sprangle}}]{esarey1994selfmod}%
  \BibitemOpen
  \bibfield  {author} {\bibinfo {author} {\bibfnamefont {E.}~\bibnamefont {Esarey}}, \bibinfo {author} {\bibfnamefont {J.}~\bibnamefont {Krall}}, \ and\ \bibinfo {author} {\bibfnamefont {P.}~\bibnamefont {Sprangle}},\ }\href {\doibase 10.1103/PhysRevLett.72.2887} {\bibfield  {journal} {\bibinfo  {journal} {Phys. Rev. Lett.}\ }\textbf {\bibinfo {volume} {72}},\ \bibinfo {pages} {2887} (\bibinfo {year} {1994})}\BibitemShut {NoStop}%
\bibitem [{\citenamefont {Nakajima}\ \emph {et~al.}(1995)\citenamefont {Nakajima}, \citenamefont {Fisher}, \citenamefont {Kawakubo}, \citenamefont {Nakanishi}, \citenamefont {Ogata}, \citenamefont {Kato}, \citenamefont {Kitagawa}, \citenamefont {Kodama}, \citenamefont {Mima}, \citenamefont {Shiraga}, \citenamefont {Suzuki}, \citenamefont {Yamakawa}, \citenamefont {Zhang}, \citenamefont {Sakawa}, \citenamefont {Shoji}, \citenamefont {Nishida}, \citenamefont {Yugami}, \citenamefont {Downer},\ and\ \citenamefont {Tajima}}]{nakajima1995selfmod}%
  \BibitemOpen
  \bibfield  {author} {\bibinfo {author} {\bibfnamefont {K.}~\bibnamefont {Nakajima}}, \bibinfo {author} {\bibfnamefont {D.}~\bibnamefont {Fisher}}, \bibinfo {author} {\bibfnamefont {T.}~\bibnamefont {Kawakubo}}, \bibinfo {author} {\bibfnamefont {H.}~\bibnamefont {Nakanishi}}, \bibinfo {author} {\bibfnamefont {A.}~\bibnamefont {Ogata}}, \bibinfo {author} {\bibfnamefont {Y.}~\bibnamefont {Kato}}, \bibinfo {author} {\bibfnamefont {Y.}~\bibnamefont {Kitagawa}}, \bibinfo {author} {\bibfnamefont {R.}~\bibnamefont {Kodama}}, \bibinfo {author} {\bibfnamefont {K.}~\bibnamefont {Mima}}, \bibinfo {author} {\bibfnamefont {H.}~\bibnamefont {Shiraga}}, \bibinfo {author} {\bibfnamefont {K.}~\bibnamefont {Suzuki}}, \bibinfo {author} {\bibfnamefont {K.}~\bibnamefont {Yamakawa}}, \bibinfo {author} {\bibfnamefont {T.}~\bibnamefont {Zhang}}, \bibinfo {author} {\bibfnamefont {Y.}~\bibnamefont {Sakawa}}, \bibinfo {author} {\bibfnamefont {T.}~\bibnamefont {Shoji}}, \bibinfo {author} {\bibfnamefont {Y.}~\bibnamefont {Nishida}},
  \bibinfo {author} {\bibfnamefont {N.}~\bibnamefont {Yugami}}, \bibinfo {author} {\bibfnamefont {M.}~\bibnamefont {Downer}}, \ and\ \bibinfo {author} {\bibfnamefont {T.}~\bibnamefont {Tajima}},\ }\href {\doibase 10.1103/PhysRevLett.74.4428} {\bibfield  {journal} {\bibinfo  {journal} {Phys. Rev. Lett.}\ }\textbf {\bibinfo {volume} {74}},\ \bibinfo {pages} {4428} (\bibinfo {year} {1995})}\BibitemShut {NoStop}%
\bibitem [{\citenamefont {Fedeli}\ \emph {et~al.}(2022)\citenamefont {Fedeli}, \citenamefont {Huebl}, \citenamefont {Boillod-Cerneux}, \citenamefont {Clark}, \citenamefont {Gott}, \citenamefont {Hillairet}, \citenamefont {Jaure}, \citenamefont {Leblanc}, \citenamefont {Lehe}, \citenamefont {Myers}, \citenamefont {Piechurski}, \citenamefont {Sato}, \citenamefont {Zaim}, \citenamefont {Zhang}, \citenamefont {Vay},\ and\ \citenamefont {Vincenti}}]{WarpX}%
  \BibitemOpen
  \bibfield  {author} {\bibinfo {author} {\bibfnamefont {L.}~\bibnamefont {Fedeli}}, \bibinfo {author} {\bibfnamefont {A.}~\bibnamefont {Huebl}}, \bibinfo {author} {\bibfnamefont {F.}~\bibnamefont {Boillod-Cerneux}}, \bibinfo {author} {\bibfnamefont {T.}~\bibnamefont {Clark}}, \bibinfo {author} {\bibfnamefont {K.}~\bibnamefont {Gott}}, \bibinfo {author} {\bibfnamefont {C.}~\bibnamefont {Hillairet}}, \bibinfo {author} {\bibfnamefont {S.}~\bibnamefont {Jaure}}, \bibinfo {author} {\bibfnamefont {A.}~\bibnamefont {Leblanc}}, \bibinfo {author} {\bibfnamefont {R.}~\bibnamefont {Lehe}}, \bibinfo {author} {\bibfnamefont {A.}~\bibnamefont {Myers}}, \bibinfo {author} {\bibfnamefont {C.}~\bibnamefont {Piechurski}}, \bibinfo {author} {\bibfnamefont {M.}~\bibnamefont {Sato}}, \bibinfo {author} {\bibfnamefont {N.}~\bibnamefont {Zaim}}, \bibinfo {author} {\bibfnamefont {W.}~\bibnamefont {Zhang}}, \bibinfo {author} {\bibfnamefont {J.}~\bibnamefont {Vay}}, \ and\ \bibinfo {author} {\bibfnamefont {H.}~\bibnamefont
  {Vincenti}},\ }in\ \href {https://doi.ieeecomputersociety.org/} {\emph {\bibinfo {booktitle} {2022 SC22: International Conference for High Performance Computing, Networking, Storage and Analysis (SC) (SC)}}}\ (\bibinfo  {publisher} {IEEE Computer Society},\ \bibinfo {address} {Los Alamitos, CA, USA},\ \bibinfo {year} {2022})\ pp.\ \bibinfo {pages} {25--36}\BibitemShut {NoStop}%
\bibitem [{\citenamefont {Picksley}\ \emph {et~al.}(2020{\natexlab{b}})\citenamefont {Picksley}, \citenamefont {Alejo}, \citenamefont {Cowley}, \citenamefont {Bourgeois}, \citenamefont {Corner}, \citenamefont {Feder}, \citenamefont {Holloway}, \citenamefont {Jones}, \citenamefont {Jonnerby}, \citenamefont {Milchberg}, \citenamefont {Reid}, \citenamefont {Ross}, \citenamefont {Walczak},\ and\ \citenamefont {Hooker}}]{picksley2020}%
  \BibitemOpen
  \bibfield  {author} {\bibinfo {author} {\bibfnamefont {A.}~\bibnamefont {Picksley}}, \bibinfo {author} {\bibfnamefont {A.}~\bibnamefont {Alejo}}, \bibinfo {author} {\bibfnamefont {J.}~\bibnamefont {Cowley}}, \bibinfo {author} {\bibfnamefont {N.}~\bibnamefont {Bourgeois}}, \bibinfo {author} {\bibfnamefont {L.}~\bibnamefont {Corner}}, \bibinfo {author} {\bibfnamefont {L.}~\bibnamefont {Feder}}, \bibinfo {author} {\bibfnamefont {J.}~\bibnamefont {Holloway}}, \bibinfo {author} {\bibfnamefont {H.}~\bibnamefont {Jones}}, \bibinfo {author} {\bibfnamefont {J.}~\bibnamefont {Jonnerby}}, \bibinfo {author} {\bibfnamefont {H.~M.}\ \bibnamefont {Milchberg}}, \bibinfo {author} {\bibfnamefont {L.~R.}\ \bibnamefont {Reid}}, \bibinfo {author} {\bibfnamefont {A.~J.}\ \bibnamefont {Ross}}, \bibinfo {author} {\bibfnamefont {R.}~\bibnamefont {Walczak}}, \ and\ \bibinfo {author} {\bibfnamefont {S.~M.}\ \bibnamefont {Hooker}},\ }\href {\doibase 10.1103/PhysRevAccelBeams.23.081303} {\bibfield  {journal} {\bibinfo  {journal} {Phys.
  Rev. Accel. Beams}\ }\textbf {\bibinfo {volume} {23}},\ \bibinfo {pages} {081303} (\bibinfo {year} {2020}{\natexlab{b}})}\BibitemShut {NoStop}%
\bibitem [{\citenamefont {Miao}\ \emph {et~al.}(2020)\citenamefont {Miao}, \citenamefont {Feder}, \citenamefont {Shrock}, \citenamefont {Goffin},\ and\ \citenamefont {Milchberg}}]{miao2020}%
  \BibitemOpen
  \bibfield  {author} {\bibinfo {author} {\bibfnamefont {B.}~\bibnamefont {Miao}}, \bibinfo {author} {\bibfnamefont {L.}~\bibnamefont {Feder}}, \bibinfo {author} {\bibfnamefont {J.~E.}\ \bibnamefont {Shrock}}, \bibinfo {author} {\bibfnamefont {A.}~\bibnamefont {Goffin}}, \ and\ \bibinfo {author} {\bibfnamefont {H.~M.}\ \bibnamefont {Milchberg}},\ }\href {\doibase 10.1103/PhysRevLett.125.074801} {\bibfield  {journal} {\bibinfo  {journal} {Phys. Rev. Lett.}\ }\textbf {\bibinfo {volume} {125}},\ \bibinfo {pages} {074801} (\bibinfo {year} {2020})}\BibitemShut {NoStop}%
\bibitem [{\citenamefont {Fournier}\ \emph {et~al.}(2011)\citenamefont {Fournier}, \citenamefont {Bonelle}, \citenamefont {Moulinec}, \citenamefont {Shang}, \citenamefont {Sunderland},\ and\ \citenamefont {Uribe}}]{fournier2011}%
  \BibitemOpen
  \bibfield  {author} {\bibinfo {author} {\bibfnamefont {Y.}~\bibnamefont {Fournier}}, \bibinfo {author} {\bibfnamefont {J.}~\bibnamefont {Bonelle}}, \bibinfo {author} {\bibfnamefont {C.}~\bibnamefont {Moulinec}}, \bibinfo {author} {\bibfnamefont {Z.}~\bibnamefont {Shang}}, \bibinfo {author} {\bibfnamefont {A.}~\bibnamefont {Sunderland}}, \ and\ \bibinfo {author} {\bibfnamefont {J.}~\bibnamefont {Uribe}},\ }\href {\doibase https://doi.org/10.1016/j.compfluid.2011.01.028} {\bibfield  {journal} {\bibinfo  {journal} {Computers \& Fluids}\ }\textbf {\bibinfo {volume} {45}},\ \bibinfo {pages} {103} (\bibinfo {year} {2011})},\ \bibinfo {note} {22nd International Conference on Parallel Computational Fluid Dynamics (ParCFD 2010)}\BibitemShut {NoStop}%
\bibitem [{\citenamefont {Archambeau}\ \emph {et~al.}(2004)\citenamefont {Archambeau}, \citenamefont {Méchitoua},\ and\ \citenamefont {Sakiz}}]{code_saturne}%
  \BibitemOpen
  \bibfield  {author} {\bibinfo {author} {\bibfnamefont {F.}~\bibnamefont {Archambeau}}, \bibinfo {author} {\bibfnamefont {N.}~\bibnamefont {Méchitoua}}, \ and\ \bibinfo {author} {\bibfnamefont {M.}~\bibnamefont {Sakiz}},\ }\href {https://hal.archives-ouvertes.fr/hal-01115371/document} {\bibfield  {journal} {\bibinfo  {journal} {International Journal on Finite Volumes}\ }\textbf {\bibinfo {volume} {1}} (\bibinfo {year} {2004})}\BibitemShut {NoStop}%
\bibitem [{\citenamefont {McLachlan}\ and\ \citenamefont {Peel}(2000)}]{gmmBook}%
  \BibitemOpen
  \bibfield  {author} {\bibinfo {author} {\bibfnamefont {G.~J.}\ \bibnamefont {McLachlan}}\ and\ \bibinfo {author} {\bibfnamefont {D.}~\bibnamefont {Peel}},\ }\href@noop {} {\emph {\bibinfo {title} {Finite Mixture Models}}},\ Wiley Series in Probability and Statistics\ (\bibinfo  {publisher} {John Wiley \& Sons, Inc},\ \bibinfo {year} {2000})\BibitemShut {NoStop}%
\bibitem [{\citenamefont {Akaike}(1998)}]{AIC}%
  \BibitemOpen
  \bibfield  {author} {\bibinfo {author} {\bibfnamefont {H.}~\bibnamefont {Akaike}},\ }\enquote {\bibinfo {title} {Information theory and an extension of the maximum likelihood principle},}\ in\ \href {\doibase 10.1007/978-1-4612-1694-0_15} {\emph {\bibinfo {booktitle} {Selected Papers of Hirotugu Akaike}}},\ \bibinfo {editor} {edited by\ \bibinfo {editor} {\bibfnamefont {E.}~\bibnamefont {Parzen}}, \bibinfo {editor} {\bibfnamefont {K.}~\bibnamefont {Tanabe}}, \ and\ \bibinfo {editor} {\bibfnamefont {G.}~\bibnamefont {Kitagawa}}}\ (\bibinfo  {publisher} {Springer New York},\ \bibinfo {address} {New York, NY},\ \bibinfo {year} {1998})\ pp.\ \bibinfo {pages} {199--213}\BibitemShut {NoStop}%
\bibitem [{\citenamefont {van~de Wetering}\ \emph {et~al.}(2023)\citenamefont {van~de Wetering}, \citenamefont {Hooker},\ and\ \citenamefont {Walczak}}]{vandewetering2023stability}%
  \BibitemOpen
  \bibfield  {author} {\bibinfo {author} {\bibfnamefont {J.~J.}\ \bibnamefont {van~de Wetering}}, \bibinfo {author} {\bibfnamefont {S.~M.}\ \bibnamefont {Hooker}}, \ and\ \bibinfo {author} {\bibfnamefont {R.}~\bibnamefont {Walczak}},\ }\href {\doibase 10.1103/PhysRevE.108.015204} {\bibfield  {journal} {\bibinfo  {journal} {Phys. Rev. E}\ }\textbf {\bibinfo {volume} {108}},\ \bibinfo {pages} {015204} (\bibinfo {year} {2023})}\BibitemShut {NoStop}%
\bibitem [{\citenamefont {Ammosov}\ \emph {et~al.}(1986)\citenamefont {Ammosov}, \citenamefont {Delone},\ and\ \citenamefont {Krainov}}]{ADK}%
  \BibitemOpen
  \bibfield  {author} {\bibinfo {author} {\bibfnamefont {M.~V.}\ \bibnamefont {Ammosov}}, \bibinfo {author} {\bibfnamefont {N.~B.}\ \bibnamefont {Delone}}, \ and\ \bibinfo {author} {\bibfnamefont {V.~P.}\ \bibnamefont {Krainov}},\ }\href@noop {} {\bibfield  {journal} {\bibinfo  {journal} {SovietPhys. JETP}\ }\textbf {\bibinfo {volume} {64}},\ \bibinfo {pages} {1191} (\bibinfo {year} {1986})}\BibitemShut {NoStop}%
\bibitem [{\citenamefont {Andreev}\ \emph {et~al.}(1998)\citenamefont {Andreev}, \citenamefont {Chizhonkov}, \citenamefont {Frolov},\ and\ \citenamefont {Gorbunov}}]{ANDREEV1998469}%
  \BibitemOpen
  \bibfield  {author} {\bibinfo {author} {\bibfnamefont {N.}~\bibnamefont {Andreev}}, \bibinfo {author} {\bibfnamefont {E.}~\bibnamefont {Chizhonkov}}, \bibinfo {author} {\bibfnamefont {A.}~\bibnamefont {Frolov}}, \ and\ \bibinfo {author} {\bibfnamefont {L.}~\bibnamefont {Gorbunov}},\ }\href {\doibase https://doi.org/10.1016/S0168-9002(98)00181-8} {\bibfield  {journal} {\bibinfo  {journal} {Nuclear Instruments and Methods in Physics Research Section A: Accelerators, Spectrometers, Detectors and Associated Equipment}\ }\textbf {\bibinfo {volume} {410}},\ \bibinfo {pages} {469} (\bibinfo {year} {1998})}\BibitemShut {NoStop}%
\end{thebibliography}%

\appendix
\section{Supplemental Material}

\subsection{Channel-formation} \label{subsec:Channel formation}
The HOFI plasma channel was generated by  the channel-forming beam, reflected by a holed mirror (hole radius \SI{15}{mm}) to enable coupling of the channel-forming and drive beams into the gas target co-axially. The beam energy remaining at the interaction point was measured using an energy meter to be $\sim$\SI{100}{mJ}. The channel-forming beam was focused to a longitudinally-extended focus by an UVFS axicon lens with \ang{3.6} base angle, resulting in \ang{1.6} approach angle of the rays to the axis. The axicon had a hole in the centre of radius \SI{13}{mm}, and was placed \SI{560(2)}{mm} from the front plate of the gas target. The axicon focus had a Bessel-function transverse profile with a measured full-width at half-maximum (FWHM) spot size of \SI{9.8(0.1)}{\micro \metre}.

\subsection{Pulse train generation} 
\label{subsec:pulseTrainGen}

\begin{figure}[t]
    \centering
    \includegraphics[width = 0.8\linewidth]{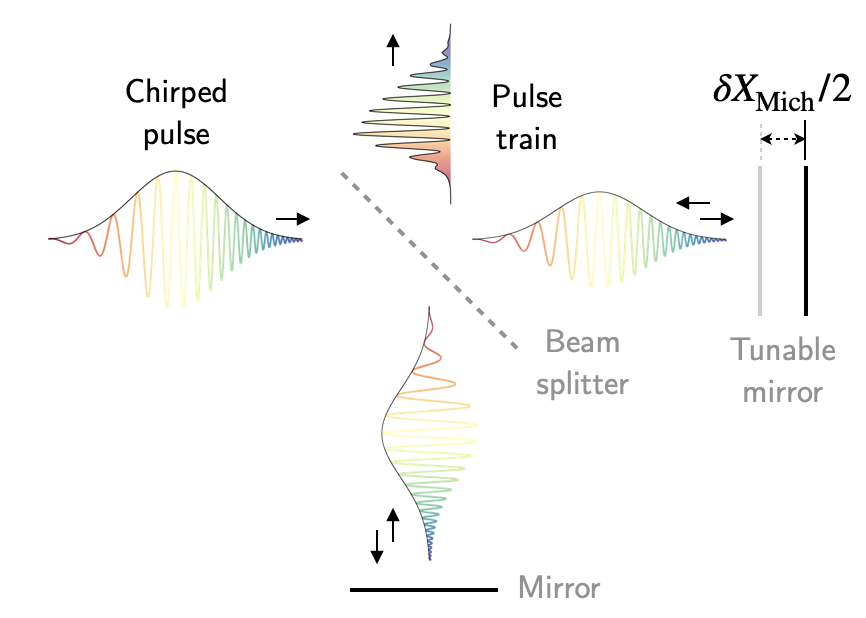}
    \caption{Schematic of the method used to generate a pulse train for the experiment. The chirped laser pulse (left) passes through a Michelson interferometer, with the two arms separated in optical path length by $\delta X_\mathrm{Mich}$. The interference pattern following the Michelson forms a pulse train, with an example intensity temporal profile shown at the top.}
    \label{fig:SFig1}
\end{figure}

Pulse trains were generated using the same method used in our earlier work~ \cite{cowley2017,Shalloo2016}. A Michelson interferometer was installed after the final amplifier in the laser chain, before the grating compressor. A flat (unwedged) compensator plate was used in the `reflected' arm of the Michelson to account for the one extra pass of the beamsplitter glass made by the `transmitted' arm. The mirror on the transmitted arm was placed on a motorised stage for remote control of the Michelson arm optical path difference, $\delta X_\text{Mich}$. It was also possible to bypass the Michelson in the laser chain, resulting in a smooth, unmodulated pulse of the same FWHM pulse duration of \SI{1}{\pico \second}.

The compressor was set for partial compression, to leave a spectral chirp on the laser beam. This resulted in unwanted higher-order phase terms being present in the partially-compressed pulse. These third and fourth order delay (TOD and FOD) spectral terms affect the uniformity of the pulse spacing in the pulse trains. An acousto-optic programmable dispersive filter (AOPDF or `Dazzler') was used to reduce the magnitude of the TOD and FOD terms in order to generate a pulse train with uniform spacing.

We can extract a measure of uniformity for these pulse trains, $U$, defined as $U \equiv 1 - (\tau_\mathrm{max} - \tau_\mathrm{min})/2\tau_{avg}$ where $\tau_\mathrm{max}$ is the maximum spacing, $\tau_\mathrm{min}$ is the minimum spacing and $\tau_\mathrm{avg}$ is the average spacing between the pulses. A TOD value of \SI{1e4}{fs^3} was found to correspond to a uniformity measure of $U \sim 0.9$ on each train. As the number of pulses in the train was relatively low, on the order of $10$, the width of the resonance was expected to be wide enough not to be significantly affected by variations in uniformity on the scale of $10 \%$, hence the uniformity of the pulse trains achieved was deemed to be sufficient.\par

\subsection{Pulse train measurement}

The temporal intensity profiles of the pulse trains were deduced from single shot autocorrelator (SSA) measurements, in conjunction with information about the settings of the grating compressor and measurements of the input spectrum, as in our earlier work~\cite{Shalloo2016, cowley2017}.\par
The measurements of the pulse train were conducted as follows. With the laser in a pulsed alignment mode, the pulse train was intercepted prior to its focus and recollimated by a lens with focal length $f = -\SI{1000}{\milli \metre}$. The beam was then sent out of the vacuum chamber via a thin optical window, and redirected to the SSA diagnostic. This consisted of a 50/50 beamsplitter, where one arm was directed via a delay stage and the other via a mirror such that the two beams crossed at an angle \ang{4} to each other. At the crossing point, a non-linear crystal (BBO Type 1, \ang{38.8} cut angle) was placed for second harmonic generation, with the generated \SI{400}{nm} light leaving the crystal perpendicular to the plane of the crystal. The unconverted \SI{800}{nm} light was dumped. A lens was used to image the blue light in the plane of the crystal on to a CCD. To increase the efficiency of blue light generation, a pair of cylindrical lenses ($f=\SI{30}{\milli \metre}$) were placed in the path of the two crossing beams, one before and one after the crystal, to focus the \SI{800}{nm} beams in the plane perpendicular to their crossing angle.\par

\begin{figure}[t]
	\centering
	\includegraphics[width = \linewidth]{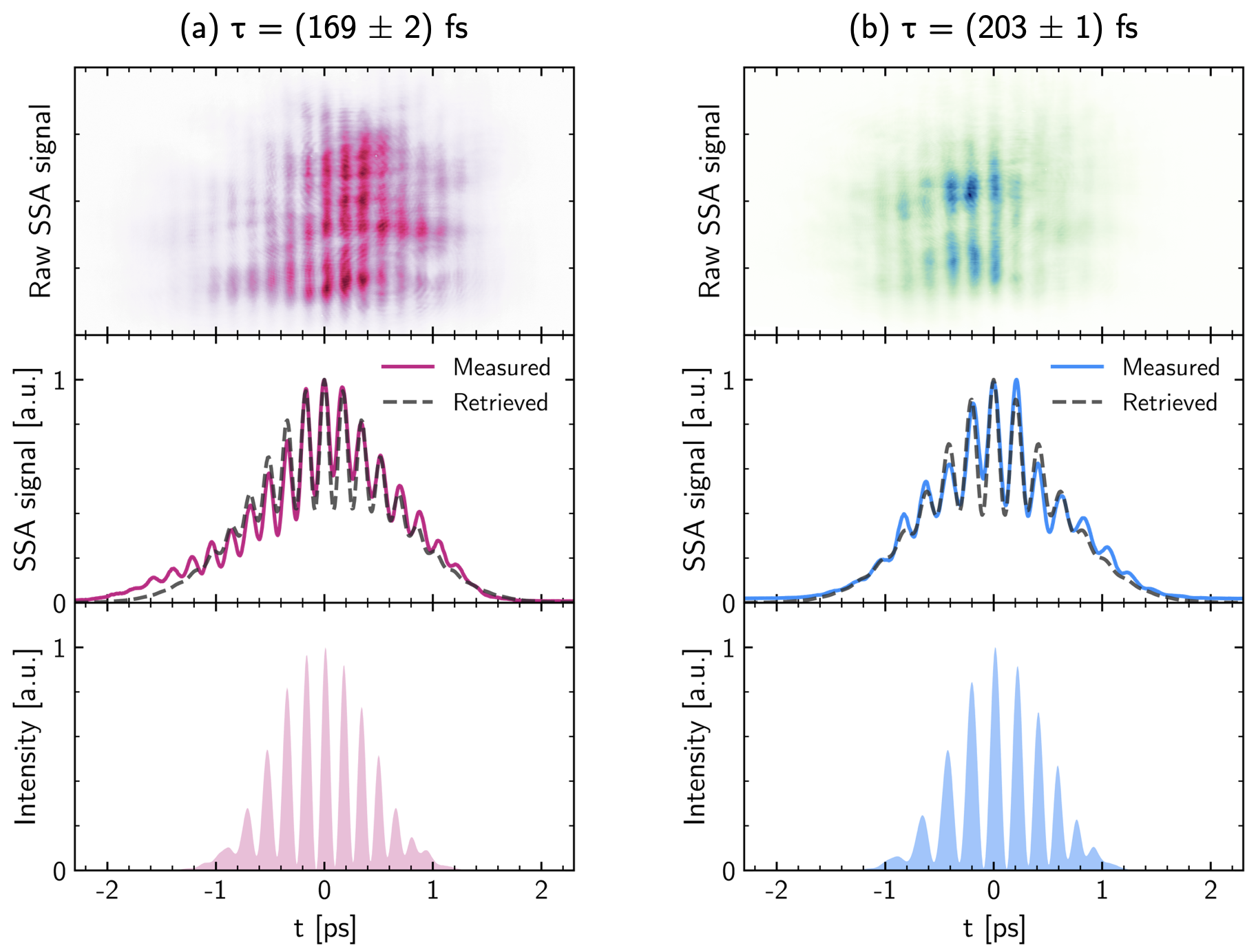}
	\caption{Example of the raw SSA signal (top), measured and retrieved SSA profiles (middle), and deduced temporal intensity profile (bottom); (a) for the case of \SI{170}{\femto \second} pulse spacing; (b) for \SI{200}{\femto \second} pulse spacing.}
	\label{fig:SFig2}
\end{figure}

\subsection{Pulse train retrieval}

The compressor grating separation and Michelson spacing were set to generate a pulse train with the desired pulse spacing, and its spectrum was measured. As described above, a controlled amount of TOD was introduced by an AOPDF in order to ensure that the pulse spacing within the train was uniform. First, a model of the compressor grating was used to estimate the TOD term expected for the grating separation. This gave an estimate of the TOD value that needed to be applied with the AOPDF. The uniformity was then optimised by scanning the TOD value about the estimate from the compressor model and analysing the fringe visibility of each pair of peaks in the autocorrelation signal, $V = (I_\mathrm{max} - I_\mathrm{min})/(I_\mathrm{max}+I_\mathrm{min})$ where $I_\mathrm{max}$ and $I_\mathrm{min}$ were the maximum and minimum intensities of each peak. For a pulse train with equal pulse spacing, the autocorrelation peaks from each pair of pulses would appear at precisely the same position on the $x$-axis of the crystal. If the pulse spacing varies between pulses, the peaks in the autocorrelation function would be slightly offset from each other in $x$, resulting in a smearing of the peaks, and hence maximising $V$ maximizes the uniformity of the pulse trains. 

Assuming the TOD and FOD terms were zero after applying this method, the only remaining parameter to be determined was the average spacing of the pulses in the train $\tau = \phi^{(2)} 2 \pi c/\delta X_\mathrm{Mich}$ where $\phi^{(2)}$ is the group delay dispersion (second order phase). The value of $\delta X_\mathrm{Mich}$ was first estimated from the spectrum of the pulse train, by fitting the modulated spectrum to the spectrum of the fully-compressed pulse multiplied by a modulation function with a period of $\Omega = 2 \pi c/\delta X_\mathrm{Mich}$. The value of $\phi^{(2)}$ was estimated from a model of the grating compressor. These estimates for $\delta X_\mathrm{Mich}$, $\phi^{(2)}$ and TOD were used to give the starting points and bounds on the retrieved pulse train properties (e.g. the retrieved value for $\delta X_\mathrm{Mich}$ was bounded to be within $10\%$ of the estimated value, and the value for $\phi^{(2)}$ was bounded to be within \SI{3000}{fs^2} of the estimated value). The retrieval method progressed via a numerical optimisation algorithm to find the values for $\delta X_\mathrm{Mich}$, $\phi^{(2)}$ and TOD that would generate the measured auto-correlation profile, while using the spectrum of the fully-compressed pulse as a constraint. This method enabled the retrieval of $\delta X_\mathrm{Mich}$ and $\phi^{(2)}$ with an uncertainty $<10\%$. 

The calculated values for the spectral phase terms along with the measured power spectrum amplitude give a full description of the pulse train in the spectral domain, and hence the temporal profile could be calculated via a Fourier transform. Figure \ref{fig:SFig2} shows the raw SSA signals and temporal intensity profiles retrieved from the autocorrelation for the two pulse trains used in the experiment.

\subsection{Drive beam focus}
The spot size of the multi-pulse drive beam at focus was calculated using the $D4\sigma$ method to be \qtyproduct[product-units = single]{41.5(2.7) x 49.4(2.0)}{\micro \metre^2}, along the minor and major axes of the ellipse respectively. The vacuum Rayleigh range was found to be $z_R =$ \SI{7.9 \pm 0.7}{\milli \metre} from a fit of the measured spot size as a function of longitudinal position to $w(z) = w_0 \sqrt{1+(z/z_R)^2}$.

The spatial jitter of the drive beam focus was measured to be \qtyproduct[product-units = single]{25.1 x 37.2}{\micro \metre^2} RMS, which is on the order of a spot size. The jitter in the channel-forming beam focus position was much smaller than that of the drive beam, measured to be \qtyproduct[product-units = single]{5.2 x 2.4}{\micro \metre^2} RMS, owing to the smaller effective $f$-number of the axicon focus ($f_{\#} = 1/(2\theta) = 11.5$) compared to the drive beam focus ($f_{\#} = 40$). 

\subsection{On-shot input spectrum measurement} \label{subsec:LA3 diagnostics}

\begin{figure}[t]
    \centering
    \includegraphics[width = 0.7\linewidth]{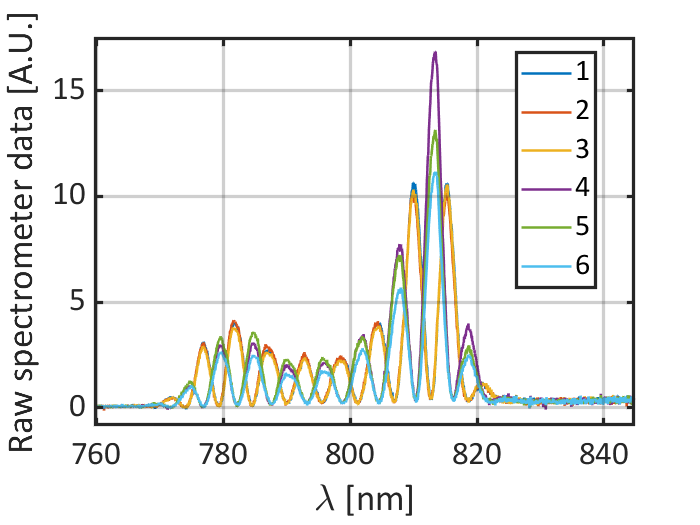}
    \caption{Raw data of the pulse-train input spectrum measured by a spectrometer in the laser area for a series of $6$ successive shots, demonstrating the shot-to-shot variation in the position of the spectral peaks. Note that the envelope is not the true spectral intensity envelope of the pulse train as the measurements here have not been corrected for the non-uniform white-light response of the measurement system.}
    \label{fig:SFig3}
\end{figure}

The spectrum of the modulated drive beam was measured directly after the the pulse-train-generating Michelson interferometer. It was found that the spectral peaks in the drive beam shifted within the spectral envelope of the beam from shot to shot (see Fig.~\ref{fig:SFig3}). This spectral jitter was believed to arise from small fluctuations in the optical path lengths in the Michelson arms. \par

In the analysis of the spectral data described in this study, the shot-to-shot spectral shifts on the input beam were accounted for with the following procedure. First, a reference spectrum for the dataset was selected and the peaks in the spectrum identified. Here, a dataset means a series of data taken within a few hour time period, with fixed settings for the pulse-generating Michelson and compressor. Then, on each shot, the peak positions in the input spectrum of that shot were compared to the positions of the reference spectrum to give a shift value for each peak, $\delta \lambda$. The mean of the peaks shift values was calculated, $\overline{\delta \lambda}$. The output spectrum, $f_\mathrm{raw}(\lambda)$ for that shot was then translated as $f(\lambda) = f_\mathrm{raw}(\lambda + \overline{\delta \lambda})$. Accounting for the shifts in the input spectrum in this way enabled the features in the transmitted spectra to be more directly compared.

\subsection{Gas target}\label{subsec:gas target}

The cell-jet hybrid gas target~\cite{Aniculaesei2018, Picksley2023} used in this experiment consisted of a cylindrical chamber of length \SI{110}{\milli \metre}, closed on each end by a plate with a \SI{6}{\milli \metre} diameter circular hole in the centre, to enable the laser pulses to enter and leave. The back plate was mounted on a motorized plunger that allowed the length of the gas target, $L$, to be varied.  The gas inlet was designed in a cone-shape to disperse the gas uniformly in the target.

Hydrogen gas was pulsed into the target via a solenoid valve, opened before the arrival of the laser pulses to ensure that the laser-plasma interaction occurred in a steady-state gas condition. 

\subsubsection*{Longitudinal density profile}

\begin{figure}[t]
     \centering
    \includegraphics[width = \linewidth]{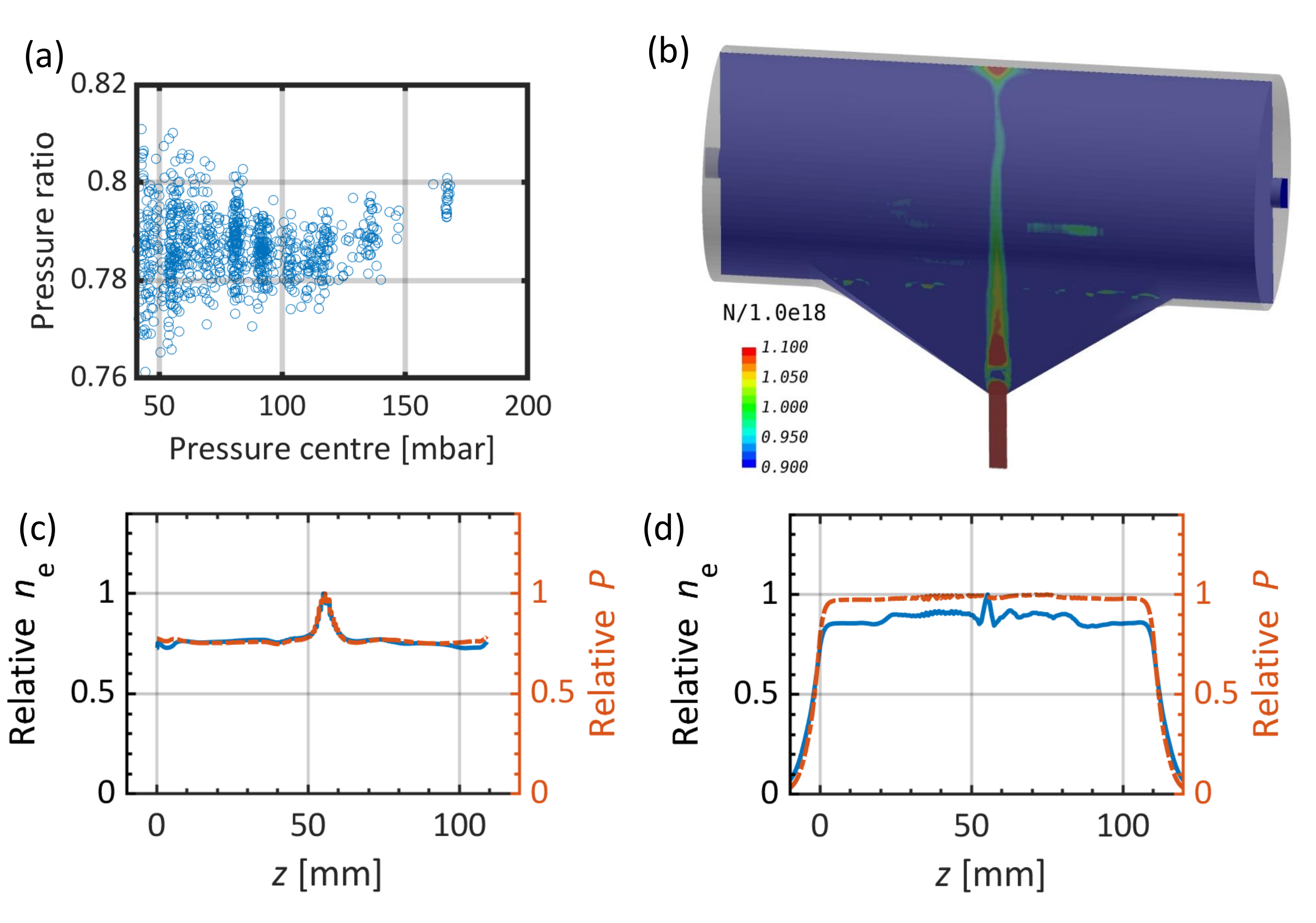}
    \caption{Gas pressure and density profiles inside the gas target. (a) shows the experimentally measured ratio of the pressure measured by the front transducer to that measured by the centre transducer. (b) shows the central slice of the 3D gas density distribution inside the target calculated using \texttt{code\_saturne}. (c) and (d) show the gas pressure and density profile lineouts extracted from the 3D fluid simulations for; (c) the top edge of the target (transducer axis); and (d) the centre line of the target (laser axis).}
    \label{fig:SFig4}
\end{figure}

The longitudinal plasma density profile depended predominantly on the longitudinal profile of the gas density inside the gas target at the moment of ionisation. During the experiment, the gas pressure was measured at two locations by pressure transducers connected to the top edge of the gas target, one near the entrance ($z=$ \SI{29}{\milli \metre} from the front pinhole) and one near the centre ($z=$ \SI{59}{\milli \metre}). Measured pressures in the target were in the range \SIrange{10}{150}{\milli \bar} and had an uncertainty of \SI{3}{\milli bar}. The ratio of these two readings as a function of pressure is shown in Fig.~\ref{fig:SFig4}(a). The mean ratio of these data is $0.79$, suggesting that the centre pressure was consistently higher than that at the front by $\sim 20\%$. \par

To gain a better understanding of the full gas profile inside the gas target, calculations were performed using the computational fluid dynamic code \texttt{code\_saturne}~\cite{fournier2011, code_saturne}. Figure \ref{fig:SFig4}(b) shows the calculated gas density distribution for the case of \SI{1}{\bar} backing pressure, in steady-state conditions. The simulations show that the gas density directly above the inlet position was greater than in the surrounding gas. In simulation, the ratio of the central pressure to the pressure at the position of the front transducer measured on the \textit{top edge} of the target shown in Fig.~\ref{fig:SFig4}(c) was found to be $0.77$, close to that found in the experiment. However, the pressure ratio between the same positions \textit{along the laser axis}, shown in Fig.~\ref{fig:SFig4}(d), was found to be $0.98$, indicating a uniform longitudinal density profile. This was confirmed via separate plasma fluorescence measurements of the longitudinal gas pressure profile along the axis of laser propagation, performed at the Oxford Plasma Accelerator Laboratory~\cite{Picksley2023}. These measurements indicated relative gas pressure fluctuations of \SI{4.1}{\%} RMS along the length of the target, consistent with the simulated pressure profile along its central axis [Fig.~\ref{fig:SFig4}(d)]. \par

\section{Selecting well-guided shots} \label{subsec:good_shots}

Due to the significant transverse jitter of the drive beam focus position ($\sim$ \SI{30}{\micro \metre}), there were a number of shots for which the drive beam was not coupled into the plasma channel. To distinguish the signal in the data corresponding to well-guided shots, a selection procedure was employed to filter the data before the analysis. The images from the exit mode diagnostic were used to define the selection criteria. When the drive beam was well-guided, the image showed a well-defined spot or group of spots. The images were analysed to give the number of spots, the spot sizes and the average pixel counts within each spot. When the beam was not guided, the image consisted of a low-amplitude, speckled light distribution. The analysis of the images for the non-guided shots identified many low-signal spot regions due to the non-uniform light distribution in the image. For the shot to be considered well-guided, the image was required to satisfy the following criteria;
\begin{enumerate}
    \item The image must contain fewer than 4 spots;
    \item All spots identified in the image must have a spot size smaller than some threshold value for that dataset;
    \item The average pixel count for each identified spot must be above some threshold value.
\end{enumerate}

\begin{figure}[t]
    \centering
    \includegraphics[width = \linewidth]{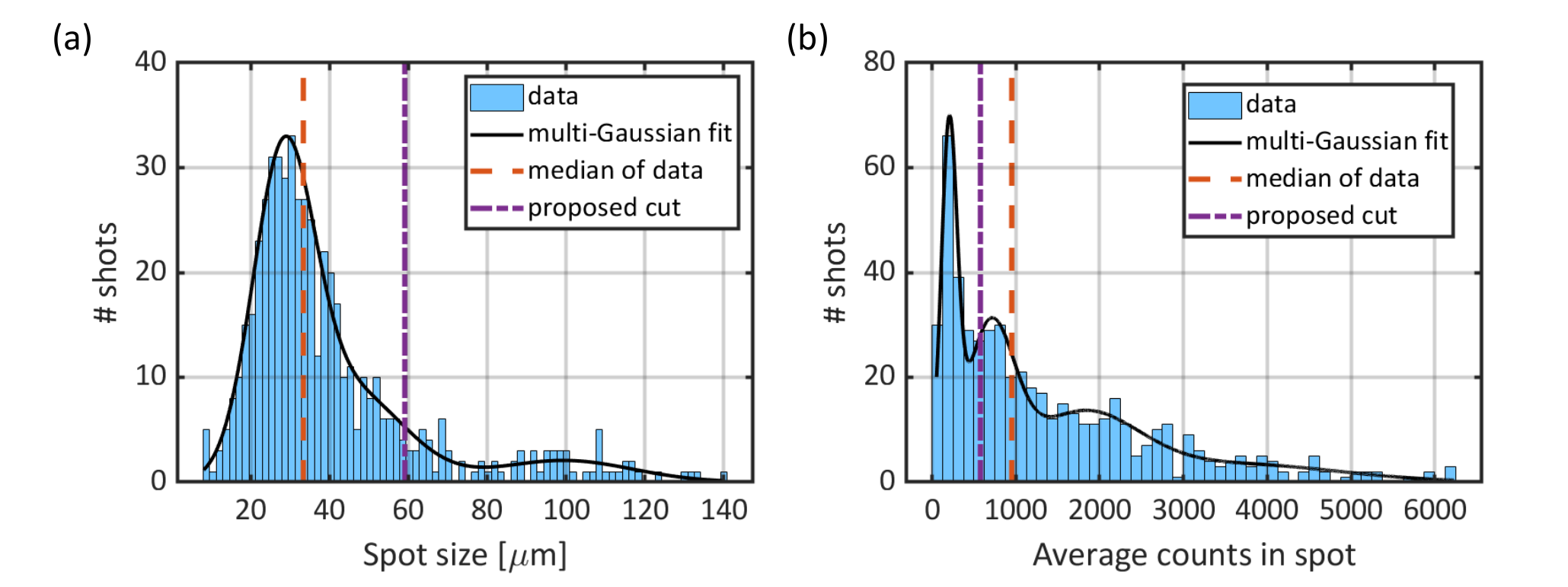}
    \caption{Histograms showing; (a) the spot size; and (b) the average pixel count in the spot, extracted from the analysis of the exit mode diagnostic images for the dataset with pulse spacing \SI{170}{fs} and gas target length \SI{110}{mm}. When multiple spots were detected in the image, the spot with the largest spot size is plotted. The black curve indicates the fitted Gaussian mixture model. The red dashed line shows the median value of the data in each case. The purple dashed line shows the threshold values calculated by the method described in the text. All data with spot size greater than the spot size threshold value and average counts lower than the average counts threshold value were discarded.}
    \label{fig:SFig5}
\end{figure}

The threshold values for spot size and average pixel counts were determined for each dataset separately. One dataset refers to a series of shots taken with the same input beam properties over a few-hour time period. This accounted for any differences in the alignment of the imaging system to the forward diagnostics between datasets. Histograms of the spot size and average fluence distributions for a representative dataset are shown in Figure \ref{fig:SFig5}. A Gaussian mixture model (GMM)~\cite{gmmBook} was fit to the data in each histogram to identify peaks in the distribution. The GMM returns a sum of $n$ Gaussian components for each distribution, with the $i^{th}$ peak having a mean $\overline{x}_{i}$ and standard deviation $\sigma_{i}$. The number of peaks in the fitted GMM was allowed to vary between 1 and 6. The optimal value of $n$ was selected using the Akaike information criterion~\cite{AIC}, a metric used to evaluate GMMs that includes both a goodness-of-fit measure and a preference for fewer components over larger models to avoid over-fitting. 

When considering the spot size, the Gaussian peak corresponding to the set with the largest mean spot size was taken as the set of failed shots. Therefore, the threshold spot size was calculated as $\overline{x}_{S,i=n-1} + 3\times \sigma_{S,i=n-1}$, where $n$ is the number of fitted components. When considering the average counts within the guided spot, the Gaussian peak in the GMM corresponding to the lowest mean pixel counts was taken as the set of failed shots. In this case, the threshold value was set at $\overline{x}_{C,i=1} + \sigma_{C,i=1}$. This procedure removed between $25\%$ and $50\%$ of the total shots from each dataset. 

\begin{figure}[t]
    \centering
    \includegraphics[width= 0.9\linewidth]{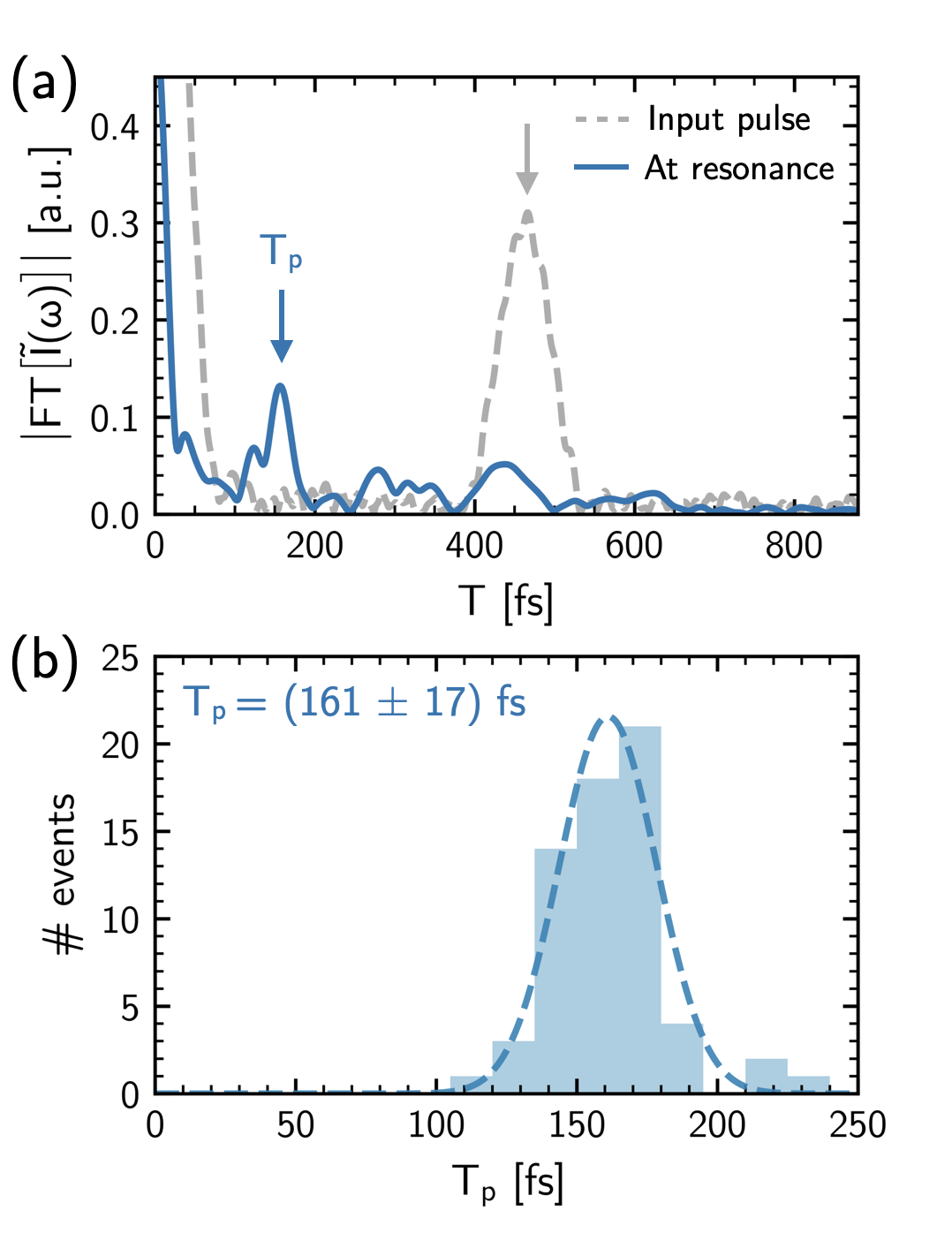}
    \caption{Procedure for the on-axis density calibration using signatures in the spectrum of the transmitted pulse train, for an example dataset with laser pulse spacing \SI{170}{fs} and gas target length \SI{110}{mm}. (a) shows the absolute value of the Fourier transform of the spectral intensity, $|FT[\tilde{I}(\omega)]|$ of the input pulse train (grey, dashed) and a transmitted pulse train at resonance (blue, solid). A strong peak in the input spectrum is observed at $\delta X_\mathrm{mich}/c \sim$ \SI{450}{fs} and a new peak in the resonant spectrum appears at $T_{\mathrm{p}} = 2\pi / \omega_p$. (b) Extracted value of $T_{\mathrm{p}}$ for 80 shots around the resonant pressure, indicating an average peak position of $T_{\mathrm{p}} = \SI{161(17)}{fs}$.}
    \label{fig:SFig6}
\end{figure}

\section{On-axis plasma density calibration}\label{subsec:densityCalibration}

Previous work has demonstrated that the on-axis density of a HOFI plasma channel is roughly linearly proportional to the initial background gas density, with a proportionality factor, $\Xi$, of between $5$ and $10$ \cite{picksley2020, feder2020} depending on the delay between the channel-forming and drive pulses. In this experiment, features in the spectrum of the transmitted pulse trains could be used to extract the value of this proportionality factor and allow conversion between pressure and on-axis plasma density of the plasma channels.

The spectral intensity of the input pulse train is modulated at $\Omega = 2\pi c/\delta X_\mathrm{Mich}$. Therefore, a Fourier transform of the spectral intensity, $FT[\tilde{I}(\omega)]$, has a dominant peak at $2 \pi/\Omega = \delta X_\mathrm{Mich}/c$. Figure~\ref{fig:SFig6}(a) shows an example case with pulse spacing \SI{170}{fs}, which exhibits a peak at $\delta X_\mathrm{Mich} / c \approx \SI{450}{fs}$. Note that this peak does \emph{not} occur at the pulse spacing $\tau = \SI{170}{fs}$, since the pulse spacing is determined by $\delta X_\mathrm{Mich}$ and the GDD of the chirped input pulse. Near resonance, the high-amplitude wakefield spectrally modulates the laser pulse train, generating side-bands in the spectrum at $\pm m \omega_p$. In the linear regime, these side-bands have the same spectral shape as the input laser spectrum but have a reduced amplitude and are offset by multiples of $\omega_p$. In the case where only the highest amplitude side-bands are visible (with $m=1$), we would expect the spectral intensity of the transmitted pulse to consist of the input pulse train spectrum, and two copies of the input pulse train spectrum, shifted by $\omega_p$ on either side of the central frequency of the pulse train spectrum $\omega_0$. Therefore, the Fourier transform of the spectral intensity at resonance is expected to consist of the peak at $\delta X_\mathrm{Mich}/c$, as well as an additional peak at $2\pi / \omega_p = T_{\mathrm{p}}$, as demonstrated in Fig.~\ref{fig:SFig6}(a).

The value of the proportionality constant $\Xi$ was extracted from the data using a fit to the observed peak positions, $T_{\mathrm{p}} = \SI{161(17)}{fs}$, for 80 shots around the observed resonant pressure [see Fig.~\ref{fig:SFig6}(b)]. To perform this analysis the spectral intensity below $\lambda_{\text{min}} \approx \SI{820}{nm}$ was set to zero, to focus on the new red-shifted light only, $\tilde{I}_{\text{red}}(\omega)$, and the position of the highest amplitude peak in $|FT[\tilde{I}_{\text{red}}(\omega)]|$ was selected as $T_{\mathrm{p}}$. This is related to the on-axis density via $T_{\mathrm{p}} = 2\pi / \omega_p$ and hence implies $n_{\mathrm{e,0}} = \SI{4.8(0.5)e17}{cm^{-3}}$. The factor to convert between the initial electron density $n_\mathrm{e}$ and the on-axis density $n_\mathrm{e,0}$ was then calculated from the fit to the data as $\Xi = n_\mathrm{e}/n_\mathrm{e,0} =$ \SI{8.1(0.6)}{}. This value is within the range expected for HOFI plasma channels \cite{picksley2020,feder2020}. 

When the gas target was set shorter than full length, the gas reading on the pressure transducers was reduced for the same backing pressures. This may have been due to a partial covering of the transducer outlet by the gas target plunger and/or a change in the flow dynamics. The on-axis density calibration factor was re-calculated for the shorter gas target length used, $L = \SI{70}{mm}$, to account for any differences in the pressure transducer reading. The calibration factor in this case was $\Xi =\SI{6.2(0.5)}{}$.

\section{Guiding Joule-scale pulse trains over ten centimetres}\label{sec:guidingResults}

In this section, additional measurements pertaining to the demonstration of guiding of pulse trains in long plasma channels are presented. Unless stated otherwise, the pulse trains used for this work had a pulse spacing of $\tau = \SI{170}{\femto \second}$. 

\subsection*{Drive beam input mode analysis}

\begin{figure}[t]
	\centering
	\includegraphics[width = \linewidth]{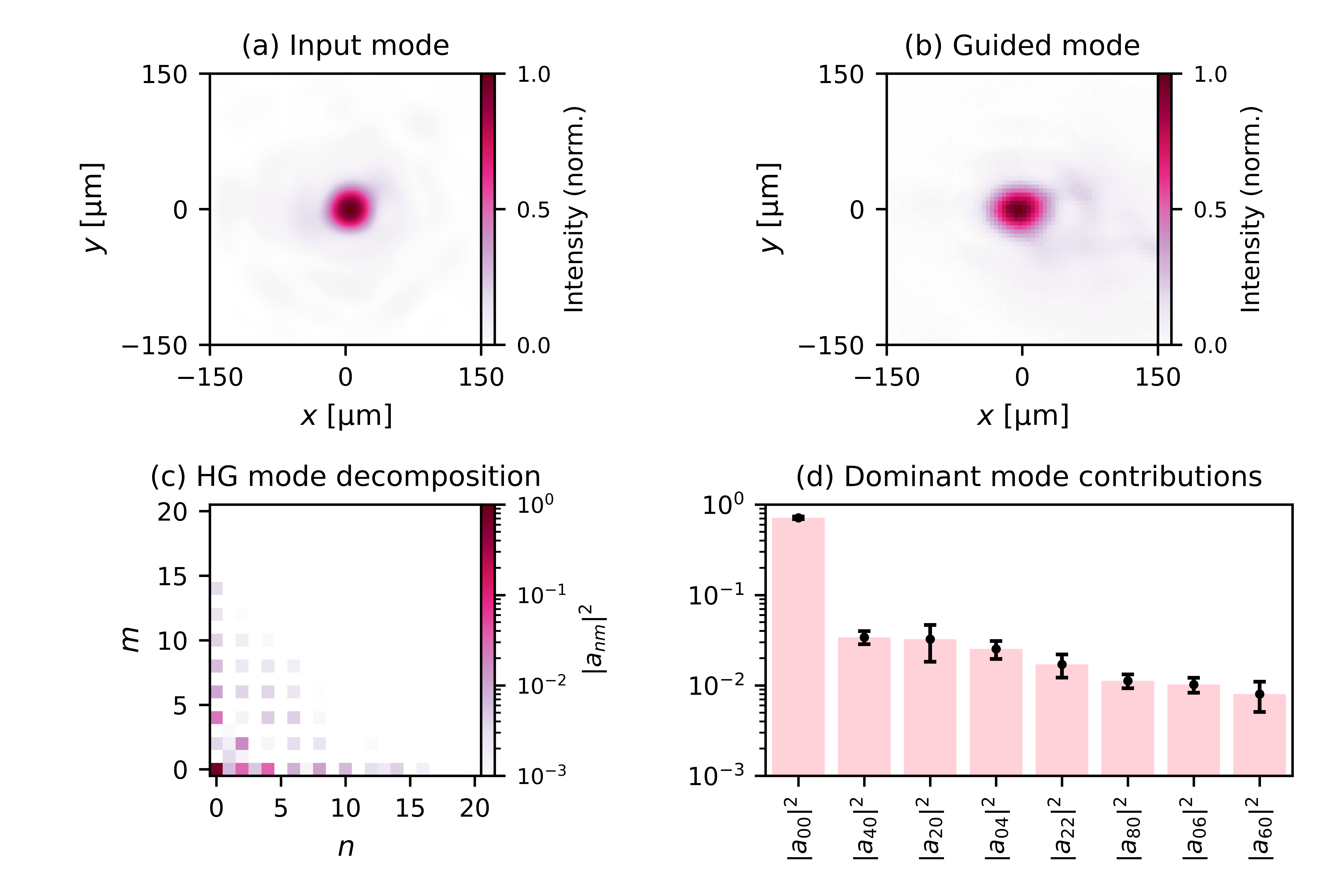}
	\caption{Comparison between examples of the (a) input mode and (b) guided mode. (c) Average value of the coefficients $|a_{nm}|^2$ for overlap integrals of 50 input modes with Hermite-Gaussian (HG) modes of spot size $w_0 = \SI{40(3)}{\micro m}$, as indicated by analysis of the highest throughput guided modes. (d) The dominant contribution is the lowest order mode: $|a_{00}|^2 = \SI{0.71(5)}{}$.}
	\label{fig:SFig7}
\end{figure}

It is desirable to match the input beam profile [Fig.~\ref{fig:SFig7}(a)] with the lowest-order mode of the plasma channel, as the lowest-order mode has the longest attenuation length. To estimate the fraction of the input beam energy that is coupled to the lowest-order mode of the plasma channel in a well-aligned case, an overlap integral can be calculated between the input beam intensity profile and the Hermite-Gaussian (HG) modes with a spot size matched to the channel. The transverse electron density profile of the channels were not measured in this experiment and so the modes of the channel could not be calculated directly. Therefore, to estimate the lowest-order mode of the channel for this calculation, the transverse intensity profile of the best-guided output beams were used to estimate the matched spot size: $w_0 = \SI{40(3)}{\micro m}$ [see Fig.~\ref{fig:SFig7}(b)]. This method assumes that the higher-order modes radiated away completely by the end of the channel, leaving only the lowest-order channel mode remaining. 

Fifty images of the input focus were used, and in each case the overlap integral of the image with the corresponding HG mode was calculated. HG modes with all combinations of indices $n, m$ varying between $0$ and $20$ were used. The average value of the coefficients over the $50$ events, $|a_{nm}|^2$, are shown in Fig.~\ref{fig:SFig7}(c), with the dominant mode contributions and their RMS fluctuations plotted in Fig.~\ref{fig:SFig7}(d). This analysis indicated that \SI{71(5)}{\%} of the input mode intensity was in the lowest order mode matched to the channel on-average.

This is a simplified analysis of the coupling of the input beam into the plasma channel, but represents an estimate of the theoretical maximum coupling efficiency of the input beam. In reality, a number of effects including density ramps at the channel entrance, the effects of conditioning of the channel by the drive beam, and the significant spatial jitter of the drive beam would all be expected to further impact the coupling of the beam into the channel. \par 

\subsection{Drive beam transmission as a function of plasma channel length}

\begin{figure}[!t]
	\centering
	\includegraphics[width = \linewidth]{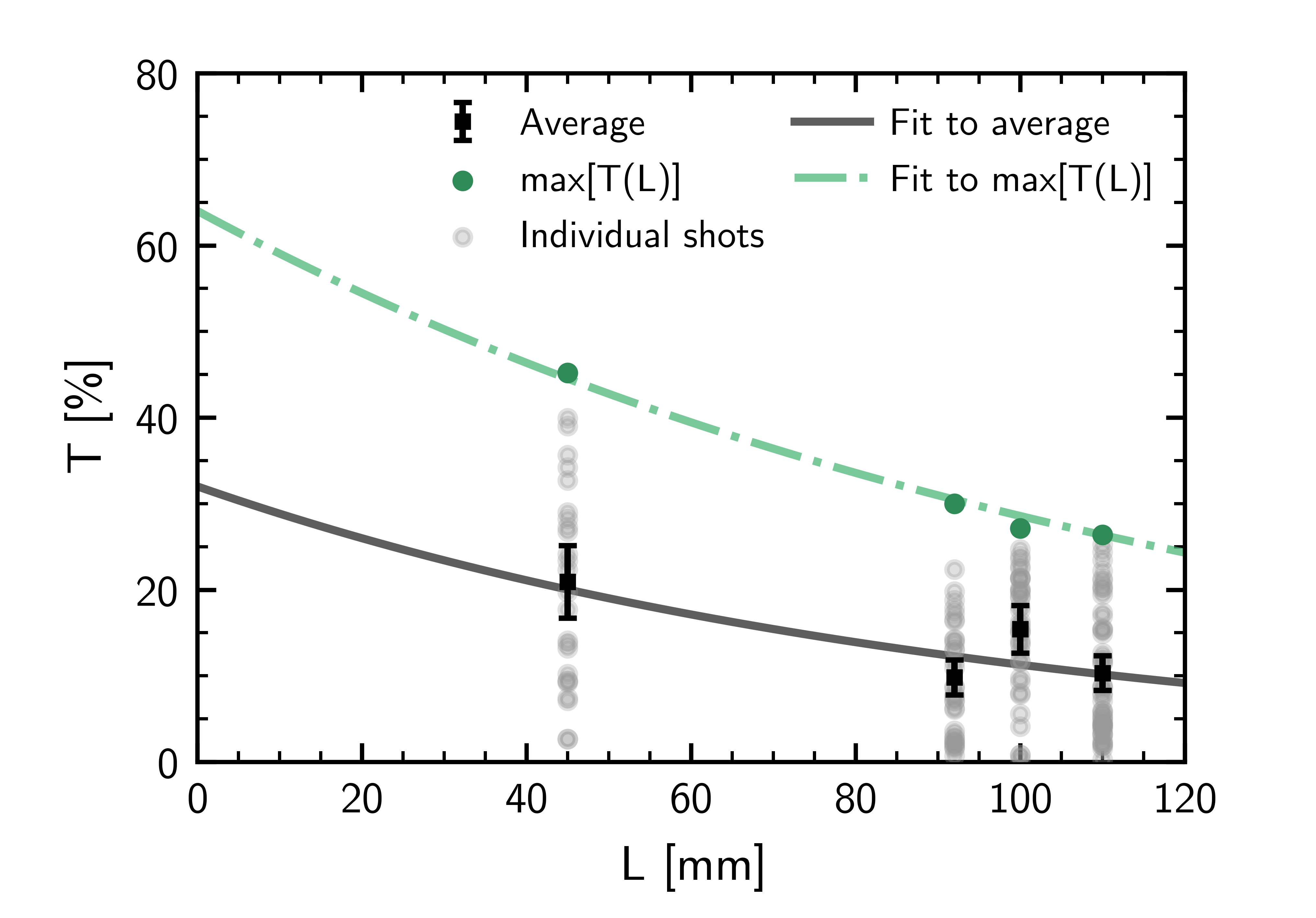}
	\caption{Measured energy transmission of the pulse train, $T$, with spacing $\tau = \SI{200}{\femto \second}$ as a function of plasma channel length, $L$, for an on-axis plasma density of $n_{\text{e,0}} = \SI{2.9e17}{cm^{-3}}$. The transmission of individual events are represented by the scattered grey data points, with the maximum at each value of $L$ plotted in green. Black, square data points show the average transmission for each channel length, with error bars representing the standard error on the mean.}
\label{fig:SFig8}
\end{figure}

To understand what fraction of the drive beam energy was coupled into the channel, the transmission of the drive beam was measured as a function of the length of the plasma channel by moving the position of the motorised plunger on the gas target. The energy in the transmitted beam was estimated by summing the pixel counts within the image of the exit modes, and comparing these to equivalent measurements of the input beam. For each shot, a correction factor for the spectral response of the CMOS sensor and transport optics within the imaging system was calculated to account for the change in the spectrum of the guided pulse arising from the plasma interaction.

The energy transmission as a function of gas target length is plotted in Fig.~\ref{fig:SFig8}, for the case of a pulse train with spacing $\tau = \SI{200}{\femto \second}$ at an on-axis density of $n_{\text{e,0}} = \SI{2.9(0.3)e17}{cm^{-3}}$. It can be seen that within each gas target length setting, there is a significant variation in energy transmission. This is mainly a result of the significant pointing jitter of the drive beam, which causes the transverse position of the input pulse to vary with respect to the axis of the plasma channel, reducing the fraction of energy that can be coupled into the guide. 

The event with highest transmission for each channel length is expected to correspond to conditions where the drive beam was most closely aligned to the channel axis, and therefore can be used to estimate the maximum coupling efficiency that was achieved in the experiment. The data for these events were fitted to an exponential decay of the form $T(z) = T_0 \exp(-z/L_\text{att})$, with $T$ the energy transmission, $T_0$ the coupling efficiency at the channel entrance, $z$ the propagation length and $L_\text{att}$ the power attenuation length. The fit yields $T_0 = \SI{64(4)}{\%}$ and $L_\text{att} = \SI{124(20)}{\milli \metre}$. The fitted coupling efficiency, $T_0 = \SI{64(4)}{\%}$, can be compared to $|c_0|^2$, where $c_0$ is the calculated coupling coefficient between the transverse amplitude profile of the incident beam and that of the lowest-order mode of the channel. This was found~\cite{Supp_mat} to be $|c_0|^2 = \SI{71(5)}{\%}$. Since higher-order modes will be attenuated rapidly, the value of $T_0$ calculated by projecting $T(z)$ to the channel input will be approximately $|c_0|^2$, and the good agreement between $|c_0|^2$ and the value of $T_0$ deduced from the well-aligned shots is consistent with this picture.  A fit to the average transmission values in Fig.~\ref{fig:SFig8} gives a coupling efficiency of $T_0 = (32 \pm 13)\%$ and $L_\text{att} = \SI{96(20)}{mm}$, which takes into account the additional coupling losses due to spatial jitter of the drive beam with respect to the channel axis. 

\subsection{On-axis plasma density dependence}\label{subsec:guidingMeasurements}

\begin{figure}[t]
	\centering 
	\includegraphics[width = 0.9\linewidth]{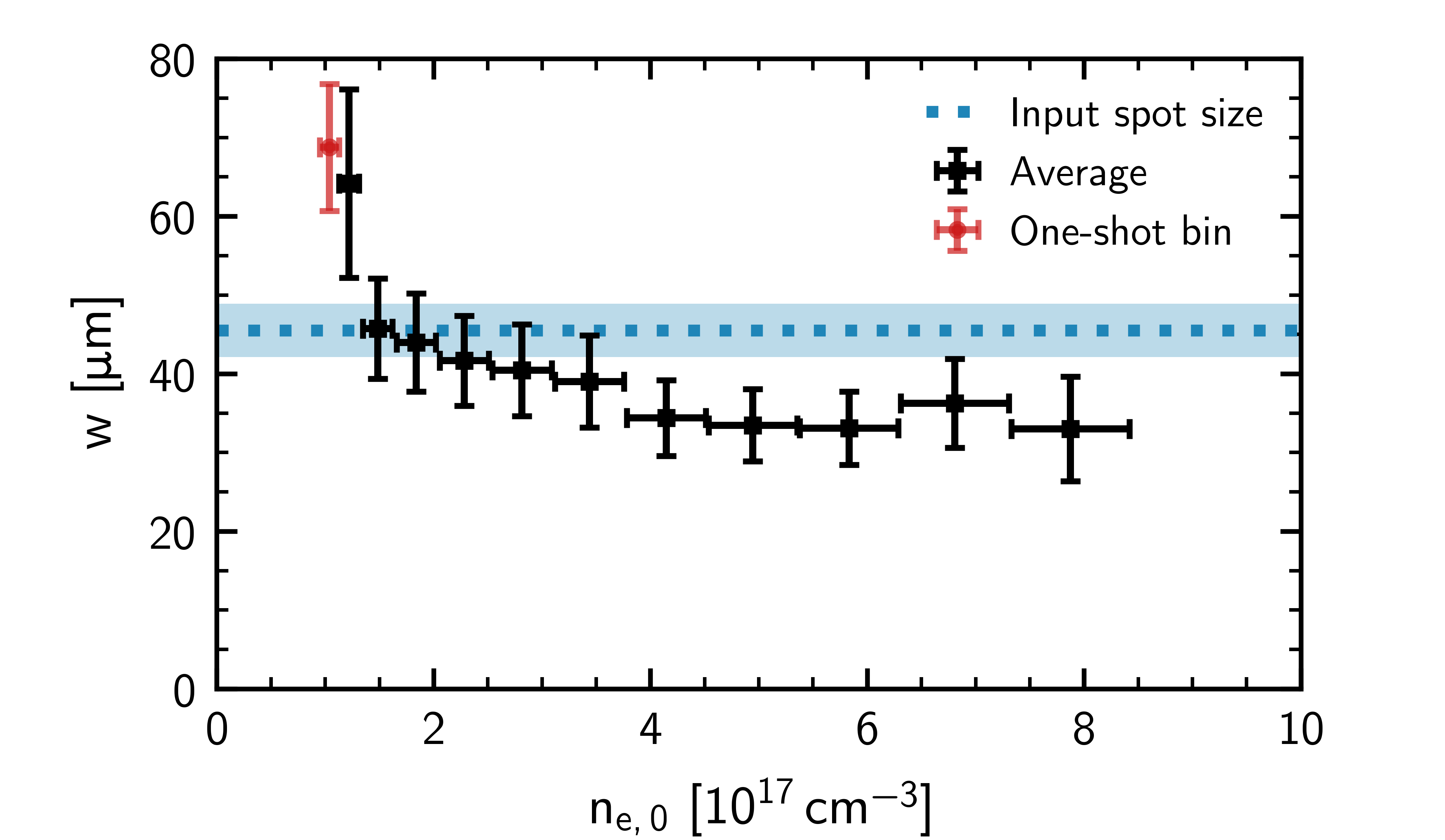}
	\caption{Measured variation of spot size of the transmitted pulse train ($E_\text{train} = \SI{2.5(0.5)}{J}$, $\tau = \SI{170}{fs}$) as a function of on-axis plasma density inside the channel ($L = \SI{110}{mm}$). The dashed blue line and shaded region shows the input spot size and RMS jitter. Red markers indicate density bins with only one shot contributing.}
	\label{fig:SFig9}
\end{figure}
 
In Fig.~\ref{fig:SFig9}, the spot size of the transmitted pulse train at the output of the plasma channel is plotted against the on-axis density in the plasma channel, over a range of densities explored in this experiment. The horizontal error bars represent the uncertainty on the measured pressure, combined with the uncertainty of the on-axis density calibration. The vertical error bars represent the standard error of the spot size calculated over all shots within each bin.

\section{2D Fluid Code}

\begin{figure}[!t]
    \centering
    \includegraphics[width= \linewidth]{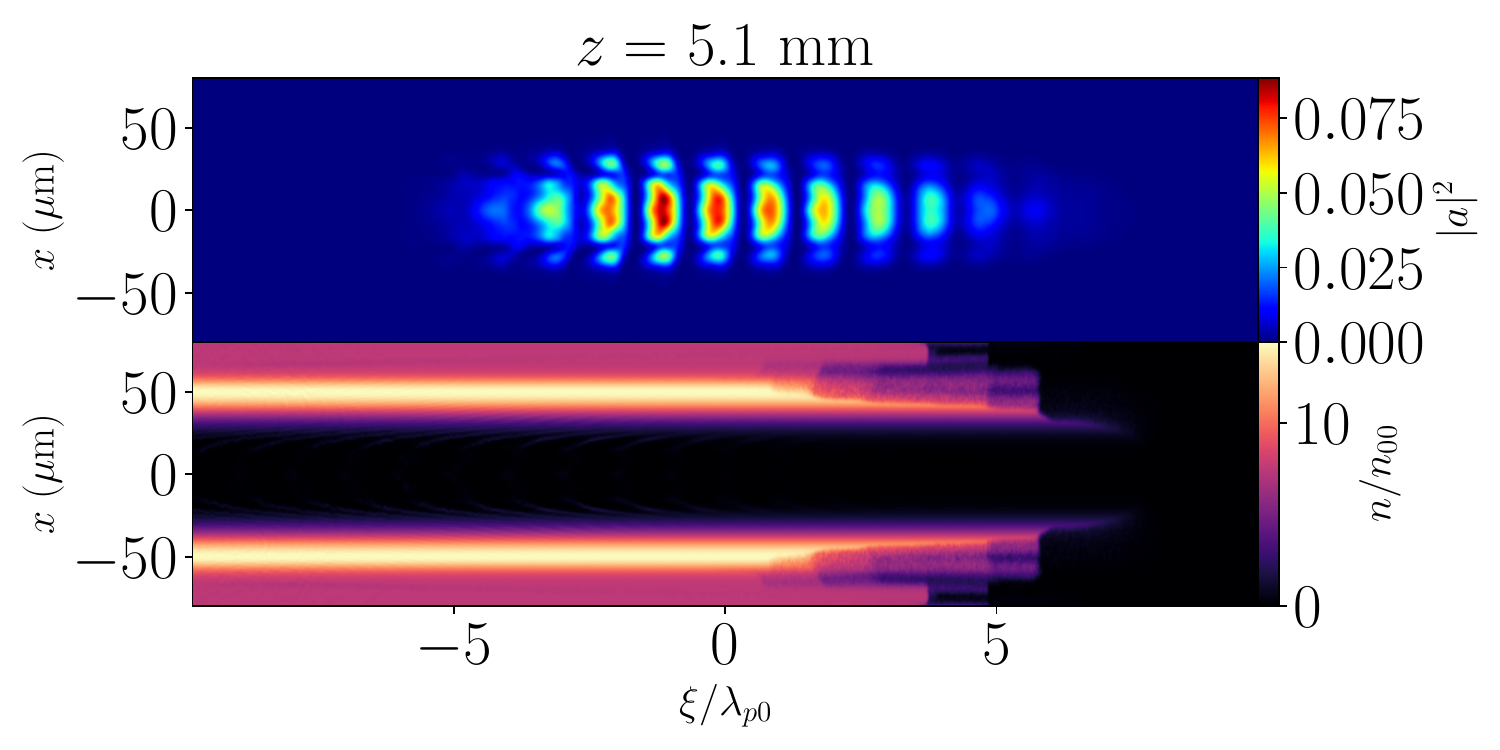}
    \caption{Results of a 2D PIC simulation of the propagation of the pulse train and its conditioning of the HOFI channel, whose profile is modelled with the parameterisation given in Eq.~(\ref{eq:CHOFI_profile}). The top panel displays the normalised intensity envelope $|a|^2$, whilst the bottom panel shows the normalised electron density $n_\text{e}/n_\text{e,0}$ including both the electrons initially ionised by the OFI process and the electrons from ionising the neutral gas collar.}
    \label{fig:SFig10}
\end{figure}

The wakefield excited by the pulse train was solved in 2D-cylindrical coordinates following the procedure outlined in Ref.~\cite{ANDREEV1998469}. For quasistatic linear wakefields driven by the normalised pulse intensity envelope $|a|^2$ in axisymmetric channels $n_\text{e}(r)$, the fluid equation for the perturbation of the normalised potential $\delta\phi=\phi-1\approx|a|^2/4-v_z/c$ and corresponding electron density perturbation $\delta n$ is given by;
\begin{align}
    \left[\left(\frac{\partial^2}{\partial\xi^2}+k_p^2(r)\right)\left(\Delta_\perp-k_p^2(r)\right)-\frac{d\ln n_\text{e}(r)}{dr}\frac{\partial^3}{\partial\xi^2\partial r}\right]\delta\phi\nonumber\\ = k_p^2(r)\left(\Delta_\perp-k_p^2(r)\right)|a|^2/4\\
    \frac{\delta n(r,\xi;|a|^2)}{n_\text{e}(r)} = k_p^{-2}(r)\left(\Delta_\perp-k_p^2(r)\right)\delta\phi+|a|^2/4
\end{align}
where $\xi=z-ct$ is the co-moving longitudinal coordinate and $k_p(r) = \omega_p(r)/c$ denotes the local plasma wavenumber. This PDE can be solved numerically for any arbitrary pulse envelope $|a|^2\ll1$ and axisymmetric plasma channel $n_\text{e}(r)$. Previous work suggests that the transverse plasma density profiles of HOFI channels are approximately parabolic \cite{feder2020}. Hence for this calculation, we assumed that the plasma channel took the form of the matched parabolic channel;
\begin{align}
    n_\text{e}(r) = n_\text{e,0} + \Delta n(r/w_0)^2
\end{align}
where $n_\text{e,0}$ is the on-axis plasma density and $\Delta n = (\pi r_ew_0^2)^{-1}$ is the channel depth parameter with $r_e$ being the classical electron radius. This channel guides the fundamental Gaussian mode $|a|^2\sim\exp(-2r^2/w_0^2)$ of spot size $w_0$. 

According to the paraxial wave equation, assuming that the ponderomotive envelope $|a|^2$ is in the fundamental Gaussian mode and remains fixed throughout the full propagation in this matched parabolic channel, and that the density perturbation is small relative to the channel depth parameter $|\delta n|\ll\Delta n$, the spectral modulation of the pulse train will be given by \cite{vandewetering2023stability};
\begin{align}
\label{eq:spectralshift}
    \frac{\Delta\omega(\xi,z_\text{end})}{\omega_0} = -z_\text{end}\frac{2c^2}{\omega_0^2w_0^2}\left\langle\frac{\partial}{\partial\xi}\frac{\delta n(r,\xi;|a|^2)}{\Delta n}\right\rangle_\perp
\end{align}
where $\Delta\omega=\omega-\omega_0$ is the shift in instantaneous frequency and $\langle(...)\rangle_\perp = (4/w_0^2)\int_0^\infty(...)\exp(-2r^2/w_0^2)rdr$ denotes the intensity-weighted transverse average. Equation~(\ref{eq:spectralshift}) was used to calculate the spectrum of the pulse train after propagation within the channel over a range of plasma densities, using the pulse envelope extracted from the pulse train retrieval process. The resulting spectra were analysed to give red-shift values, $R$, using the same method as for the experimental data. 

\section{PIC simulations of conditioning effect}

Conditioning of the neutral gas collar formed by the HOFI channel was studied using the PIC code WarpX \cite{WarpX}, which uses ADK theory \cite{ADK} to model the tunnel ionisation of neutral atoms in intense laser fields. For simplicity, a neutral gas of atomic hydrogen was assumed. The HOFI channel profile, which is comprised of the ionised electron $n_{\text{e},\text{OFI}}$ and ion $n_\text{H$^+$,OFI}$ densities from the OFI process and the density of neutral hydrogen atoms $n_\text{H}$, was parameterised in the following form
\begin{align}\label{eq:CHOFI_profile}
    &n_{\text{e},\text{OFI}}(r) = n_\text{H$^+$,OFI}(r) = n_\text{e,0}\exp\left(-r^2/d_0^2\right)\,, \\
    &n_\text{H}(r) = 
    \begin{cases}
        n_\text{max}e^{-(r-r_\text{max})^2/d_1^2}, & r\leq r_\text{max}\\
        n_\infty + \left(n_\text{max}-n_\infty\right)e^{-(r-r_\text{max})^2/d_2^2}, & r>r_\text{max}
    \end{cases}
\end{align}
where $n_\text{e,0}=\SI{4.3e17}{cm^{-3}}$, $n_\text{max}=\SI{6.2e18}{cm^{-3}}$, $n_\infty=\SI{3.2e18}{cm^{-3}}$, $d_0=\SI{45}{\micro m}$, $d_1=\SI{16}{\micro m}$, $d_2=\SI{8}{\micro m}$, and $r_\text{max}=\SI{50}{\micro m}$.

An example of the conditioning effect of the pulse train is shown in Fig.~\ref{fig:SFig10}, which shows that full conditioning of the neutral gas collar is achieved after the first few low-energy pulses have passed.

\end{document}